\documentclass[aps,prc,superscriptaddress,showpacs,amssymb,amsmath,amsfonts,floatfix,nofootinbib]{revtex4-1}
\usepackage{graphicx}
\usepackage{natbib}
\usepackage[usenames]{color}
\usepackage{epsfig}
\usepackage{epstopdf}
\usepackage{amssymb,amsmath}
\usepackage{subfig}
\usepackage{lineno,hyperref}
\usepackage{enumitem}
\usepackage{subfig}
\usepackage{array}
\usepackage{multirow}
\usepackage{amstext}
\usepackage{cleveref}
\usepackage[dvipsnames]{xcolor}

\definecolor{grey}{rgb}{0.9,0.9,0.9}
\definecolor{black}{rgb}{0,0,0}

\newcommand{\fpf}[2]{{F}_{#1}^{*}{F}_{#2}}

\hypersetup{pdfpagemode=FullScreene,pdfstartview={XYZ null null 1.1}}

\newcommand{\Tuzla}[1]
{\affiliation{University of Tuzla, Faculty of Natural Sciences and Mathematics, \\ Univerzitetska 4, 75000 Tuzla, Bosnia and Herzegovina}}
\newcommand{\Mainz}[1]
{\affiliation{Institut f\"ur Kernphysik, Johannes Gutenberg-Universit\"at Mainz, D-55099
Mainz,Germany}}
\newcommand{\Zagreb}[1]
{\affiliation{Rudjer Bo\v{s}kovi\'{c} Institute, Bijeni\v{c}ka cesta 54, P.O. Box 180, 10002 Zagreb, Croatia}}
\newcommand{\Tesla}[1]
{\affiliation{Tesla Biotech, Mandlova 7, 10002 Zagreb, Croatia}}

\begin{document}

\title{Single-Energy Partial Wave Analysis for $\pi^0$ Photoproduction on Proton with \\ Fixed-$t$ Analyticity Imposed \vspace*{0.2cm}}

\author{H.~Osmanovi\'{c}}\thanks{hedim.osmanovic@untz.ba}\Tuzla \\
\author{M.~Had\v{z}imehmedovi\'{c}}\Tuzla \\
\author{R.~Omerovi\'{c}}\Tuzla \\
\author {J.~Stahov}\Tuzla\\
\author{M.~Gorchtein}\Mainz\\
\author{V.~Kashevarov}\Mainz\\
\author{K.~Nikonov}\thanks{Now at Helmholtz-Institut f{\"u}r Strahlen- und Kernphysik, University of Bonn, Bonn, Germany}\Mainz\\
\author{M.~Ostrick}\Mainz\\
\author{L.~Tiator}\Mainz \\
\author{\vspace*{0.5cm} A.~\v{S}varc}\Zagreb,\Tesla


\date{\today}

\begin{abstract}
\vspace*{1.cm}
High precision data of the  $\gamma p \to \pi^0 p$ reaction from its
threshold up to $W=2$~GeV have been used in order to perform a
single-energy partial wave analysis with minimal model dependence. Continuity in
energy was achieved by imposing constraints from fixed-$t$
analyticity in an iterative procedure. Reaction models were only
used as starting point in the very first iteration. We demonstrate
that with this procedure partial wave amplitudes can be obtained
which show only a minimal dependence on the initial model
assumptions.

\end{abstract}

\maketitle

\section{Introduction} \label{sec:intro} 
The excitation spectra of nucleons are studied since the 1950's and provided important pieces of information during the discovery of quarks and color charge. However, despite of this long history the spectrum is still not fully established and much less understood. At high energies ($W > 2$~GeV)  various quark models and lattice QCD predict an almost exponential increase in the density of states which could so far not be confirmed experimentally (\textquotedblleft missing resonance problem\textquotedblright). In this paper we discuss the lower energy range $W < 2$~GeV, where most analyses agree on the number of states, however, with sizable uncertainties in the decay properties and even in the excitation energies. Because of their extremely short lifetime, excited nucleons appear as resonance poles in complex partial wave scattering amplitudes. The reliable and unambiguous determination of these  amplitudes is a central task which requires both, precision measurements of several spin-dependent observables and sophisticated analysis methods. Major progress was made in meson photo- and electroproduction due to the availability of high intensity polarized beams and polarized targets in combination with 4$\pi$ detector systems. In our analysis we use polarization data  obtained at ELSA, GRAAL, JLab and MAMI. Data with unprecedented quality and quantity are available, in particular for photoproduction of pions.

Theoretical single or multi-channel models were developed and used to determine the resonance parameters. These approaches are called energy-dependent (ED) analyses because the energy dependence of amplitudes is parameterized in terms of resonant and non-resonant contributions. The model parameters are estimated by fits to the data. Ideally, such models should respect fundamental constraints like analyticity, unitarity and crossing symmetry. On the other hand the computational effort should be sufficiently small to allow for detailed systematic studies during the complex fitting procedures. In practice, compromises are necessary and in general the extracted multipole amplitudes and resonance parameters vary from model to model.
A recent comparison of the prominent ED models (Bonn-Gatchina (BnGa)-\cite{BoGa}, J\"{u}lich-Bonn (J\"{u}Bo)-\cite{Juelich}, George Washington University (SAID)-\cite{SAID},  and Mainz (MAID)-\cite{MAID}) has been done in Ref.~\cite{Beck}.
%
There, for the case of the $\gamma p \to \pi^0 p$ reaction, it has been demonstrated that the model dependence can be reduced and that multipoles obtained in different analyses start to converge when all modern polarization data are taken into account.

In so called energy-independent or single-energy (SE) approaches a truncated partial wave expansion is fitted to the measured angular distributions independently at each individual energy bin without using a reaction model. At first sight this method seems to provide a direct connection between experimental data and multipole amplitudes. However, it has been demonstrated that possible ambiguities can only partially be resolved by high-quality experimental data alone \cite{Omelaenko:1981cr, Wunderlich:2017dby, Workman:2016irf}. All observables remain unchanged if at each energy and angular bin the reaction amplitudes are multiplied by an overall phase $\phi(W,\theta)$. This continuum ambiguity prevents a unique projection to multipoles.
In Ref.~\cite{Svarc2018} this has been studied in detail and it was shown that a single energy partial wave analysis is discontinuous in energy unless this phase is constrained by additional theoretical input.

In the simple case of pion production close to threshold and in the energy region of the $\Delta(1232)$ resonance where only a few partial waves contribute and unitarity provides strong constraints in the form of Watson's theorem \cite{Watson1954} model independent multipole analyses were possible \cite{Beck:1999ge,Hornidge:2012ca, Schumann:2015ypa, Markou:2018skh}. However, in all analyses at higher energies, the fits were constrained to a parametrization of the amplitudes given by a preferred reaction model (e.g. \cite{MAIDSE, SAIDSE, KENTSE}). These kind of analyses are useful in order to study the consistency of the energy dependent parametrization of a model, however, they do not provide independent multipole amplitudes.

 In Ref.~\cite{Osmanovic2018} we have developed a method to impose analyticity of the reactions amplitudes in the Mandelstam variable $s$ at a fixed value of the variable $t$ in an iterative procedure. A reaction model is only necessary as a starting point in the very first iteration. We have applied this method to the $\gamma p \to \eta p$ reaction and demonstrated that indeed single energy solutions can be obtained which do not have discontinuities in their energy dependence. Remaining ambiguities were traced back to limitations in the data base and different overall phases $\phi(W,\theta)$ of the initial reaction models. In this paper we apply the method to the  $\gamma p \to \pi^0 p$ reaction. On the one side, pion production is more complicated than $\eta$ production because both, excitations  with isospin $I = 1/2$ ($N$) and $I = 3/2$ ($\Delta$), contribute in the same $\pi^0 p$ multipoles. On the other side, much more experimental data are available, and the phases in different models are stronger constrained than in $\eta$ production.

The paper is organized as follows. In section~\ref{sec:formalism} we briefly describe the formalism. In section~\ref{sec:results} we comment on the experimental data that were used in our analysis and present the single-energy multipoles for different starting solutions. In section~\ref{sec:discussion} we compare our results with experimental data and give suggestions for further measurements in order to obtain a unique PWA. Finally, in the appendix we give basic formula for kinematics and polarization observables in different representations.

\section{Formalism}
\label{sec:formalism}

In this paper we apply the fixed-$t$ analyticity constraining method for single-energy partial wave analysis in $p(\gamma,\pi^0)p$, that we developed previously and applied to $\eta$ photoproduction on the proton~\cite{Osmanovic2018}. All details are described in our previous paper and only some important issues as the iterated fitting procedure are repeated here. The kinematics for pion photoproduction is presented in appendix A, cross sections and polarization observables used in our analysis are defined in appendix B and expressions in terms of CGLN amplitudes and helicity amplitudes are listed in appendix C.

Pseudoscalar meson photoproduction on the nucleon, as $p(\gamma,\eta)p$ and $p(\gamma,\pi^0)p$ are fully described with four invariant amplitudes, e.g. $A_i(\nu,t), i=1,\ldots,4$, where $\nu$ and $t$ are Mandelstam variables, see appendix A. For a close conection between amplitudes and observables, spin (CGLN) amplitudes $F_i(W,\theta), i=1,\ldots,4$ and helicity amplitudes $H_i(W,\theta), i=1,\ldots,4$ are defined in the meson-nucleon c.m. frame, where $W$ is the total c.m. energy and $\theta$ is the c.m. angle of the outgoing meson. While for $N(\gamma,\eta)N$ the isospin in the final $\eta N$ system is $\frac{1}{2}$ and only two channels proton and neutron exist, in pion photoproduction the total isospin is either $\frac{1}{2}$ or $\frac{3}{2}$ and four channels $p(\gamma,\pi^0)p$, $n(\gamma,\pi^0)n$, $p(\gamma,\pi^+)n$ and $n(\gamma,\pi^-)p$ are possible. In a future (ongoing) work a full isospin analysis will be performed using our novel fixed-$t$ method. Here we concentrate only on $p(\gamma,\pi^0)p$ and, therefore, we can ignore the isospin aspect here, and treat everything just like $\eta$ photoproduction with the $\pi^0$ as a `light eta meson'.

The novelty in this paper is introducing fixed-$t$ analyticity in SE PWA to ensure the continuity of PWA solutions following the method developed for $\eta$ photoproduction in Ref.~\cite{Osmanovic2018}. Namely, if one performs a SE PWA freely, one actually uses all available data on observables ${\cal O}(W_{fixed},\theta)$ at one isolated energy at different angles, without paying any attention to what is happening at neighbouring energies. However, as it has been shown in Ref.~\cite{Svarc2018}, continuum ambiguity (invariance of reaction amplitudes to the rotation with arbitrary real, energy and angle dependent phase) enables multiplicity of equivalent, but different PWA solutions at one energy with different reaction amplitude phase, so if no continuity of the phase is imposed when we move from one energy to another one, the solution is automatically discontinuous as each different phase gives different sets of partial waves. One elegant solution to this problem is imposing the analyticity at fixed-$t$, and that automatically means continuity in energy as well. To do so, first we have to obtain reaction amplitudes in the $t$ variable which describe the used data base, but which are at the same time continuous at fixed $t$. To achieve that, we have to do an amplitude reconstruction of observables not in the ${\cal O}(W_{fixed},\theta)$  form but in  the ${\cal O}(t_{fixed},W)$ form. The first step is to interpolate the existing data to predetermined fixed-$t$ values. This procedure, called a fixed-$t$ data binning, is in details described in~\cite{Osmanovic2018}. As a second step we have to fit these data with a continuous function. As we demand a minimal model dependence, instead of using theoretical models, we use the Pietarinen expansion method first applied in PWA of $\pi N$ elastic scattering data \cite{Pietarinen, Hohler84} to describe the reaction amplitudes in fixed-$t$. We start with an arbitrary solution, and find a fixed-$t$ solution in such a way that our results do not deviate much from the starting solution using penalty function techniques. We then take the obtained continuous amplitudes as an constraint in SE PWA again using penalty function techniques to impose continuity in energy. In this way we obtain an SE solution which at the same time, describes the measured data and is continuous. This solution is different from the starting solution. With the obtained results we go back to the fixed-$t$ amplitude analysis and use it as a constraint. We continue this iterative procedure as long as the result does not change much, and this typically happens after 3-4 iterations. The final result is continuous in energy. Details of the procedure are given in ref.~\cite{Hohler84, HamiltonPetersen, Osmanovic2018}.

The method consists of two separate analyses, the fixed-$t$ amplitude analysis (FT AA) and the single energy partial wave analysis (SE PWA). The two analyses are coupled in such a way that the results from FT AA are used as a constraint in SE PWA and vice versa in an iterative procedure. It has not been proven, but it is extensively tested in  $\pi N$ elastic, fixed-$t$ constrained SE PWA~\cite{Hohler84}, and since then recommended for other processes.

\begin{figure}[htb]
\begin{center}
\includegraphics[width=10cm]{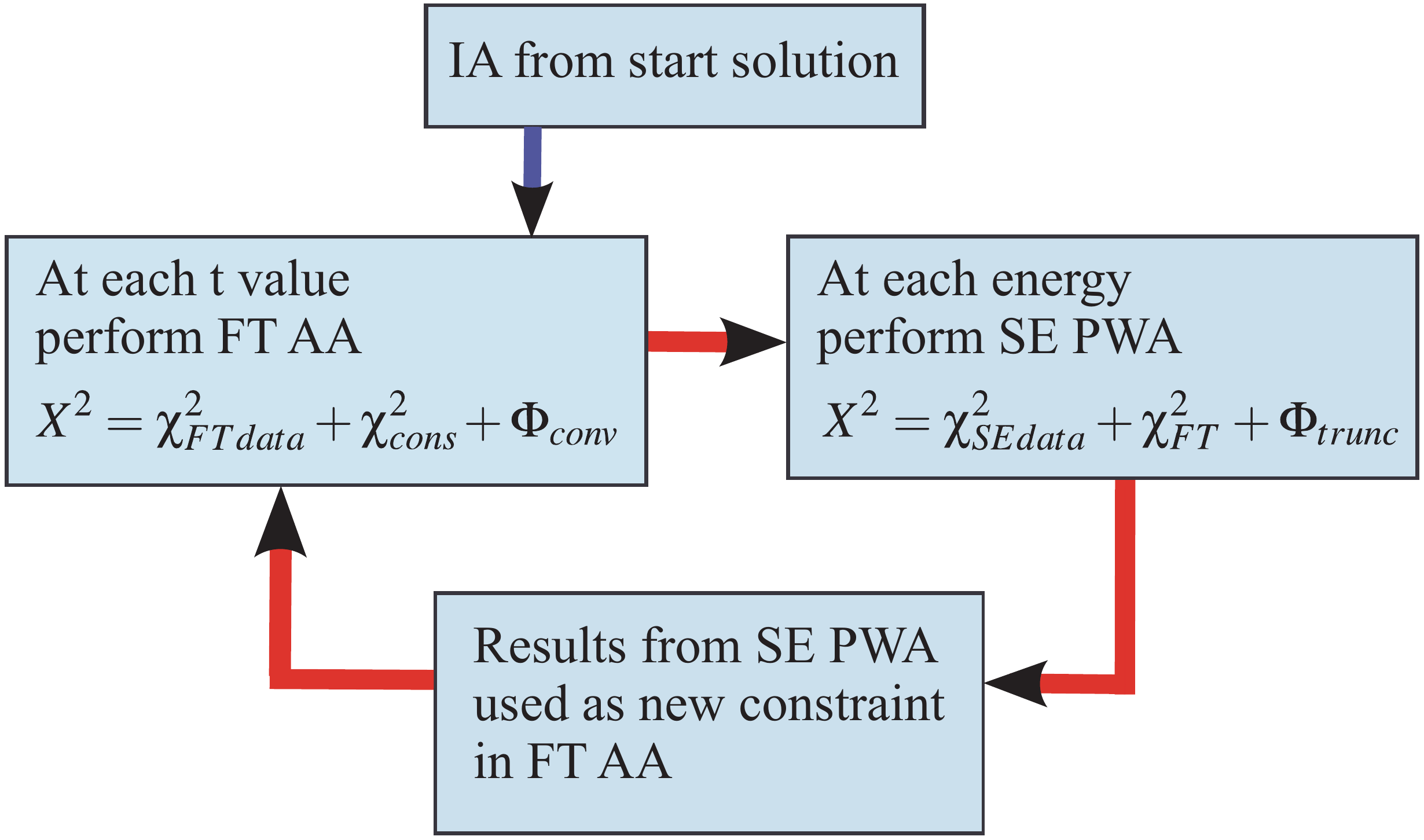}
\vspace{3mm} \caption{\label{Fig:Scheme} Iterative minimization
scheme which achieves point-to-point continuity in energy using
fixed-$t$ analyticity as a constraint. (IA: invariant amplitudes,
FT~AA: fixed-$t$ amplitude analysis, SE~PWA: single-energy partial
wave analysis) }
\end{center}
\end{figure}

\begin{itemize}[itemsep=1.ex,leftmargin=2.5cm]
\item[\bf Step 1:] Constrained FT AA is performed by minimizing the form
\begin{equation}\label{chi2_AA}
X^{2}=\chi_{FTdata}^{2}+\chi_{cons}^2+\Phi_{conv}\,,
\end{equation}

where $\chi_{cons}^2$ is a constraining term given by
\begin{eqnarray}
\chi_{cons}^{2} & = &
q_{cons}\sum_{k=1}^{4}\sum_{i=1}^{N^{E}}\left(\frac{Re\:
H_{k}(E_i)^{fit}-Re\:
H_{k}(E_i)^{cons}}{\varepsilon_{k,i}^{Re}}\right)^{2}\\
\nonumber &  &
+q_{cons}\sum_{k=1}^{4}\sum_{i=1}^{N^{E}}\left(\frac{Im\:
H_{k}(E_i)^{fit}-Im\:
H_{k}(E_i)^{cons}}{\varepsilon_{k,i}^{Im}}\right)^{2}. \nonumber
\end{eqnarray}
$H_{k}^{cons}$ are helicity amplitudes from SE PWA in the previous iteration. In a first iteration, $H_{k}^{cons}$ are calculated from the initial PWA solution (MAID, SAID, BnGa, J\"uBo). $H_{k}^{fit}$ are values of helicity amplitudes $H_{k}$ calculated from coefficients in Pietarinen's expansions, which are parameters of the fit. $N^E$ is the number of energies for a given value of $t$, and $q_{cons}$ is an adjustable weight factor. $\varepsilon_{k,i}^{Re}$ and $\varepsilon_{k,i}^{Im}$ are errors of real and imaginary parts of the corresponding helicity amplitudes. In our analysis we take $\varepsilon_{k,i}^{Re}=\varepsilon_{k,i}^{Im}=1.$

\item[\bf Step 2:] Constrained  SE PWA
is performed by minimizing the form
\begin{equation}\label{chi2-SEPWA}
X^{2}=\chi_{SEdata}^{2}+\chi_{FT}^2+\Phi_{trunc}\,,
\end{equation}
where the additional term $\chi_{FT}^2$ contains the helicity
amplitudes from the FT AA in step 1:
\begin{eqnarray}
\chi_{FT}^{2} & = &
q_{cons}\sum_{k=1}^{4}\sum_{i=1}^{N^{C}}\left(\frac{Re\:
H_{k}(\theta_{i})^{fit}-Re\:
H_{k}(\theta_{i})^{FT}}{\varepsilon_{k,i}^{Re}}\right)^{2}\\
\nonumber &  &
+q_{cons}\sum_{k=1}^{4}\sum_{i=1}^{N^{C}}\left(\frac{Im\:
H_{k}(\theta_{i})^{fit}-Im\:
H_{k}(\theta_{i})^{FT}}{\varepsilon_{k,i}^{Im}}\right)^{2}.
\nonumber
\end{eqnarray}
$N^C$ is the number of angles for a given energy $E$ and the values
$\theta_i$ are obtained for a corresponding value of $t$ using
Eq.~(\ref{fixedt}).
\item[\bf Step 3:] Use resulting multipoles obtained in step 2, and  calculate helicity amplitudes which serve
as a constraint in step 1.
\end{itemize}
$\chi^2_{FTdata}$ and $\chi^2_{SEdata}$ are standard $\chi^2$ functions calculating the weighted deviations between theory and experiment, and $\Phi_{conv}$ and $\Phi_{trunc}$ are penalty functions that are described in Ref.~\cite{Osmanovic2018}. In step 1 for the FT AA, the energy-dependent helicity amplitudes $H_{k}(E_i)^{fit}$ are parametrized with Pietarinen functions, where the expansion coefficients are the fit parameters. In step 2 for the SE PWA, the angle-dependent helicity amplitudes $H_{k}(\theta_i)^{fit}$ are parametrized with Legendre functions and multipoles, where the multipoles are the fit parameters. An iterative minimization scheme which accomplishes point-to-point continuity in energy is given in Fig.~\ref{Fig:Scheme}.

\clearpage

\section{Results}
\label{sec:results}

\subsection{\boldmath $\pi^0$ photoproduction data base}
\label{Database}
The majority of experimental data on the $\gamma p \to \pi^0 p$ reaction were obtained with energy-tagged photon beams and large solid angle electromagnetic calorimeters with a high efficiency for a detection of the 2 photons from the $\pi^0 \to \gamma \gamma$ decay. We used data from the A2 Collaboration at the Mainz Microtron MAMI \cite{Hornidge:2012ca, Adlarson2015, Leukel2001, Annand2016, Schumann:2015ypa, OttePhD, Preobrajenski2001, Linturi2015, Ahrends2005}, the CBELSA/TAPS Collaboration at  ELectron Stretcher and Accelerator ELSA \cite{Hartmann2014, Gottschall2014, Thiel2012} and the GRAAL Detector at the European Synchrotron Radiation Facility \cite{Bartalini2005}. A summary is given in Table~\ref{tab:expdata}.
\begin{table}[htb]
\centering{}\caption{\label{tab:expdata} Experimental data from A2@MAMI, DAPHNE/MAMI, CBELSA/TAPS and
GRAAL Collaborations used in our SE PWA.}
\bigskip{}
\begin{tabular}{|c|c|c|c|c|}
\hline
$Obs$  & $N$  & $W[MeV]$ & $N_{E}$  & Reference\tabularnewline
\hline
\hline
\multirow{2}{*}{$\sigma_{0}$} & $5240$  & $1075-1541$ & $262$  & A2@MAMI(2013)~\cite{Hornidge:2012ca}\tabularnewline
 & $3930$  & $1132-1895$ & $246$  & A2@MAMI(2015)~\cite{Adlarson2015}\tabularnewline
\hline
\multirow{3}{*}{$\Sigma$} & $528$  & $1074-1215$ & $54$  & A2@MAMI(2013)~\cite{Hornidge:2012ca}\tabularnewline
 & $357$  & $1150-1310$ & $21$  & A2@MAMI(2001) ~\cite{Leukel2001}\tabularnewline
 & $471$  & $1383-1922$ & $31$  & GRAAL(2005)~\cite{Bartalini2005}\tabularnewline
\hline
\multirow{2}{*}{$T$} & $469$  & $1295-1895$ & $34$  & A2@MAMI(2016)~\cite{Annand2016}\tabularnewline
 & $157$  & $1462-1620$ & $8$  & CBELSA/TAPS(2014)~\cite{Hartmann2014}\tabularnewline
$T\sigma_{0}$ & $4500$  & $1074-1291$ & $250$  & A2@MAMI(2015)~\cite{OttePhD}\tabularnewline
\hline
$P$  & $157$  & $1462-1620$ & $8$  & CBELSA/TAPS(2014)~\cite{Hartmann2014}\tabularnewline
\hline
$E\sigma_{0}$ & $139$  & $1201-1537$ & $24$  & DAPHNE/MAMI(2001)~\cite{Preobrajenski2001}\tabularnewline
\multirow{2}{*}{$E$} & $88$  & $1481-1951$ & $5$  & CBELSA/TAPS(2014)~\cite{Gottschall2014}\tabularnewline
 & $480$  & $1129-1878$ & $40$  & A2@MAMI(2015)~\cite{Linturi2015}\tabularnewline
\hline
$F$ & $469$  & $1295-1895$ & $34$  & A2@MAMI(2016)~\cite{Annand2016}\tabularnewline
$F\sigma_{0}$ & $4500$  & $1074-1291$ & $250$  & A2@MAMI(2015) ~\cite{OttePhD}\tabularnewline
\hline
\multirow{2}{*}{ $G$} & $3$  & $1232$ & $1$  & DAPHNE/MAMI(2005)~\cite{Ahrends2005}\tabularnewline
 & $318$  & $1430-1727$ & $19$  & CBELSA/TAPS(2012)~\cite{Thiel2012}\tabularnewline
\hline
$H$  & $157$  & $1462-1620$ & $8$  & CBELSA/TAPS(2014)~\cite{Hartmann2014}\tabularnewline
\hline
\end{tabular}
\end{table}

In general, there is a hierarchy of precision depending on the polarization degrees of freedom used in the experiment. The highest precision was achieved in measurements of the unpolarized differential cross section at MAMI \cite{Adlarson2015}. The statistical uncertainties are so small, that systematic uncertainties due to angular dependent detection efficiencies had to be taken into account. For all other observables the uncertainties in the angular distributions are dominated by statistics. Normalization errors (luminosity, polarization degree) are below of 5\% and are not taken into account in this analysis.

For our single energy fits we need all observables at the same values of $W = \sqrt{s}$ and for the fixed-$t$ fits at the same values of $t$. Typically this is not provided by the experiments directly. The data are given in bins of $W$ and center of mass angle $\theta_{cm}$ with bin sizes and central values varying between different data sets. Therefore some interpolation between measurements at different energies and angles is necessary. We have used a spline smoothing method~\cite{deBoor} which was similarly applied in the Karlsruhe-Helsinki analysis KH80~\cite{Hohler84} and in our previous analysis of $\eta$ production \cite{Osmanovic2018}. The uncertainties of interpolated data points are taken to be equal to the errors of nearest measured data points.
\begin{figure}[htb]
    \begin{center}
        \includegraphics[width=0.9\textwidth]{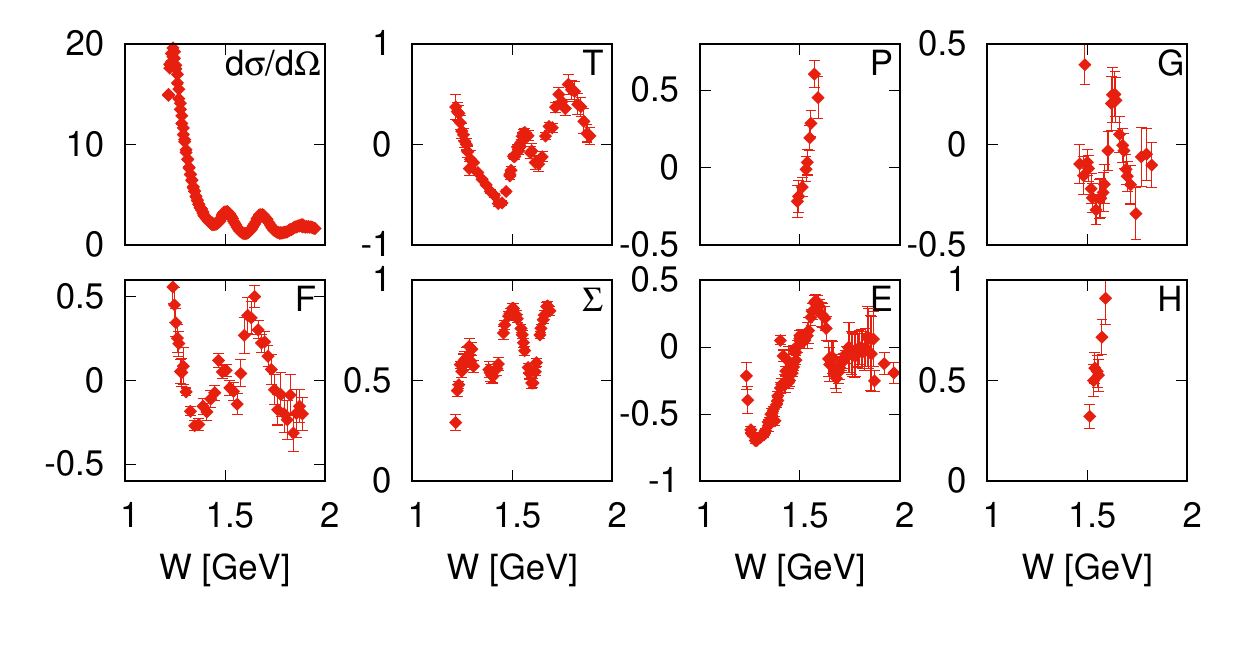}
        \includegraphics[width=0.9\textwidth]{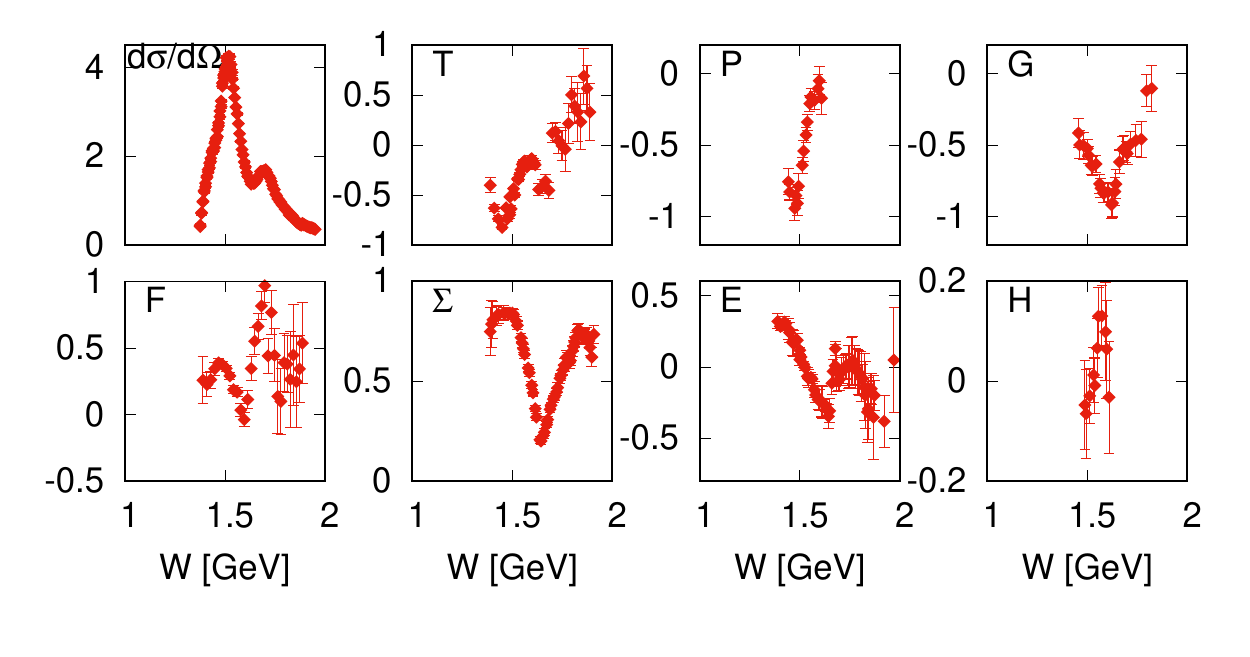}
        \vspace{3mm} \caption{\label{FigFTData} Example of our interpolated
            fixed-$t$ data base for $t = -0.2\,$GeV$^2$ (upper part)
            and  $t = -0.5\,$GeV$^2$ (lower part).}
    \end{center}
\end{figure}
Our fixed-$t$ amplitude analysis is performed at $t$ values in the range $-1.00\,$GeV$^{2}<t<-0.005\,$GeV$^{2}$ with 20 equidistant $t$ values. Examples of interpolated data points  at $t=-0.2$~GeV$^2$ and $-0.5$~GeV$^2$ are shown in Fig. \ref{FigFTData}. We note that the individual errors are not independent.

\clearpage

\subsection{\boldmath $\pi^0$  photoproduction multipoles}
\begin{figure}[h]
\begin{center}
\includegraphics[width=0.9\textwidth]{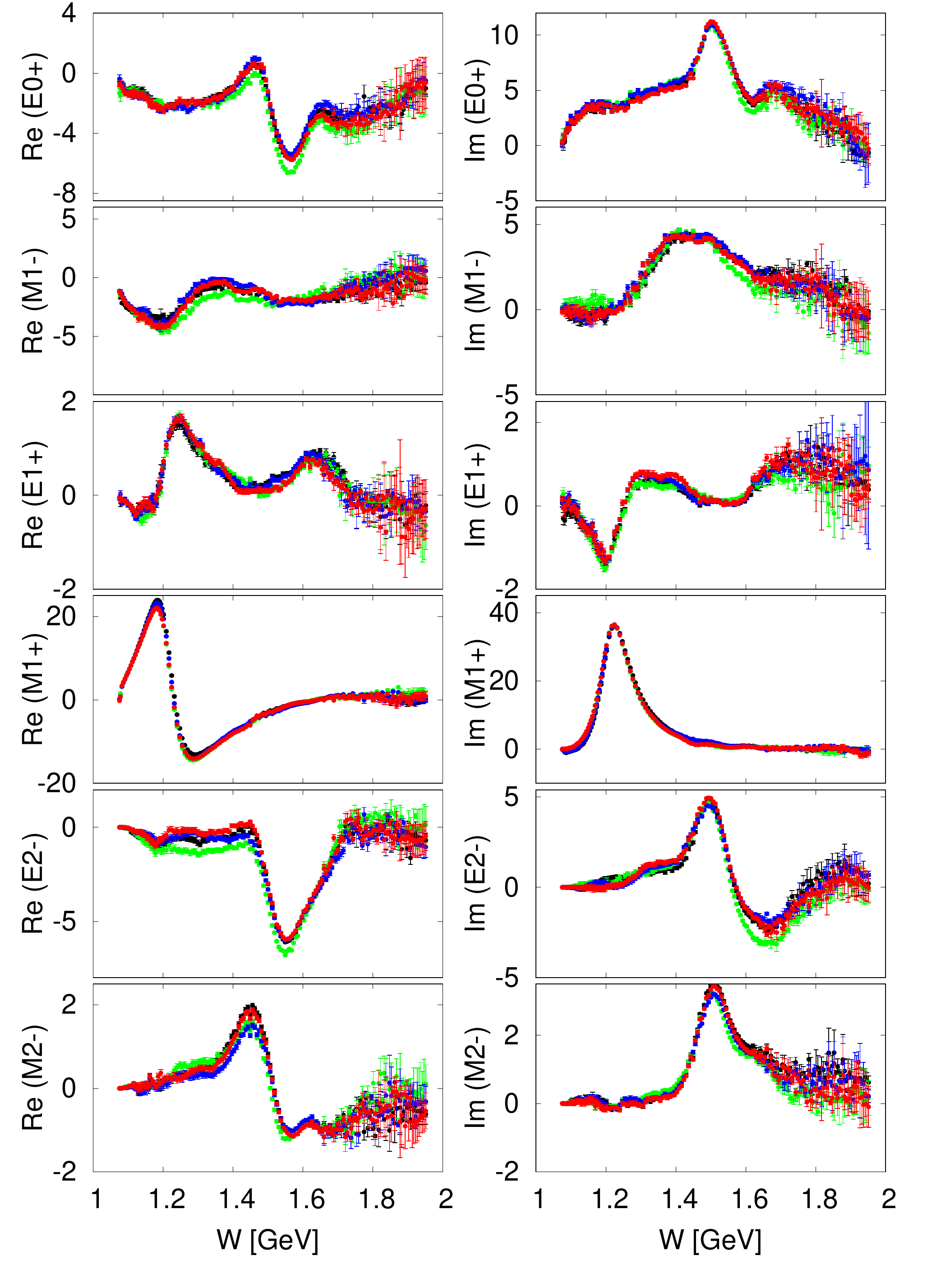}
\vspace{3mm} \caption{\label{FigMultSEpi01}
SE1, \ldots, SE4 solutions obtained using different models as initial solutions (BnGa (black), J\"uBo (blue), SAID (red) and MAID (green)), see text for further details. Multipoles are  in units of mfm.}
\end{center}
\end{figure}

\begin{figure}[h]
\begin{center}
\includegraphics[width=0.9\textwidth]{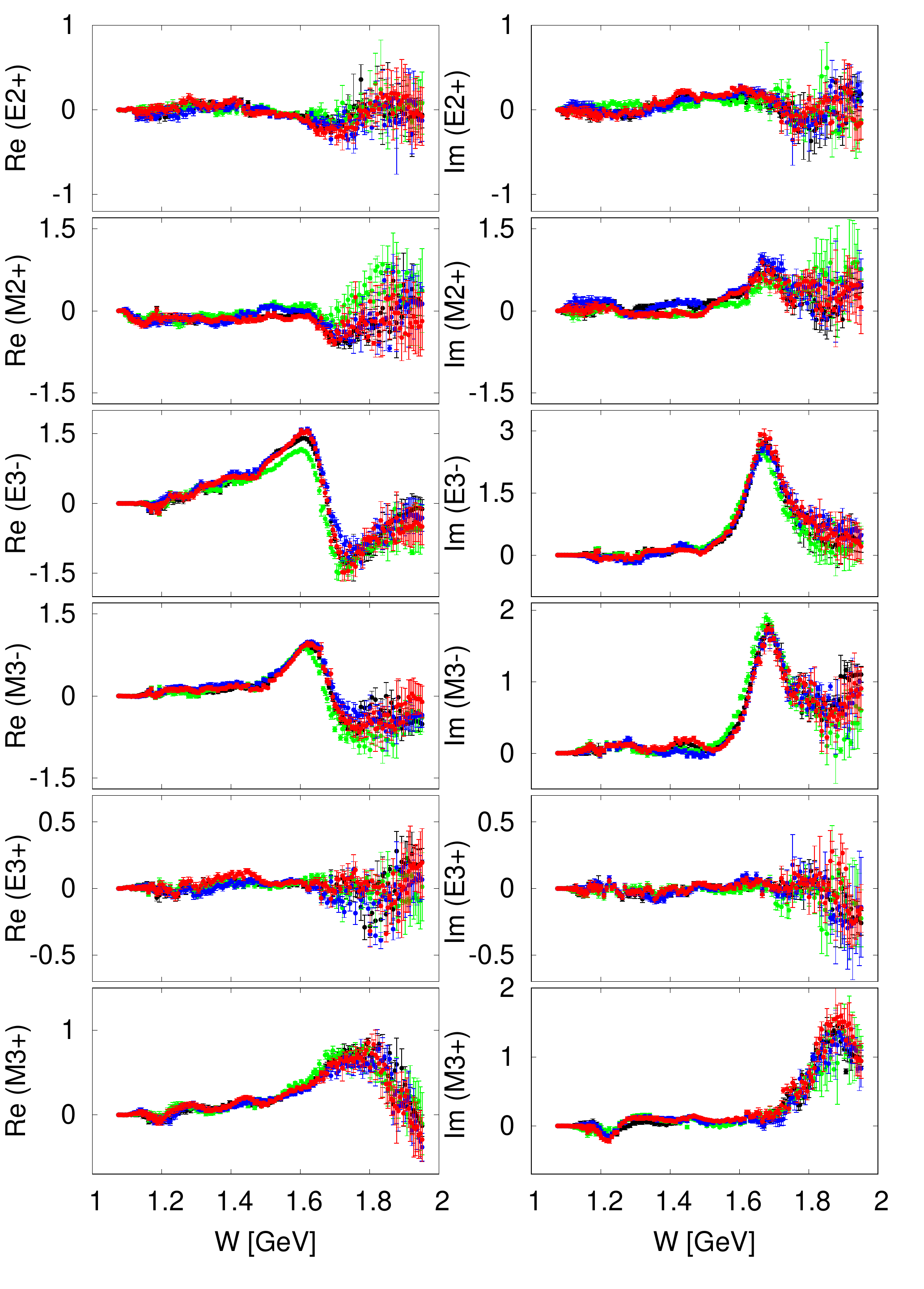}
\vspace{3mm} \caption{\label{FigMultSEpi02}
SE1, \ldots, SE4 solutions obtained using different models as initial solutions (BnGa (black), J\"uBo (blue), SAID (red) and MAID (green)), see text for further details. Multipoles are  in units of mfm.}
\end{center}
\end{figure}

Fitting was done in a standard way using MINUIT program package, and the final result is presented in Figs.~\ref{FigMultSEpi01} and \ref{FigMultSEpi02}, where four different single-energy solutions are compared. For SE1 we use as starting solution BG2014-02 from Bonn-Gatchina~\cite{BoGa}, for SE2 the J\"uBo2015-B solution from J\"ulich-Bonn~\cite{Juelich}, for SE3 the CM12 solution from SAID~\cite{SAID}, and for SE4 the MAID2007 solution from Mainz~\cite{MAID}. As can be seen from Table~\ref{tab:expdata}, at energies  \mbox{$W < 1.7$ GeV} we us as much as eight observables $\sigma_0$, $\Sigma$, $T$, $P$, $E$, $F$, $G$, $H$. However, as not all observables are taken at comparable energies, the maximum of five observables was fitted simultaneously (differential cross section + four spin observables). But the combination is not identical in each energy bin. At higher energies the number of measured observables is reduced, so at some energies we use only two spin observables in addition to the cross section.


\clearpage

\section{Discussion}
\label{sec:discussion}

\begin{figure}[htb]
\begin{center}
\includegraphics[width=0.9\textwidth]{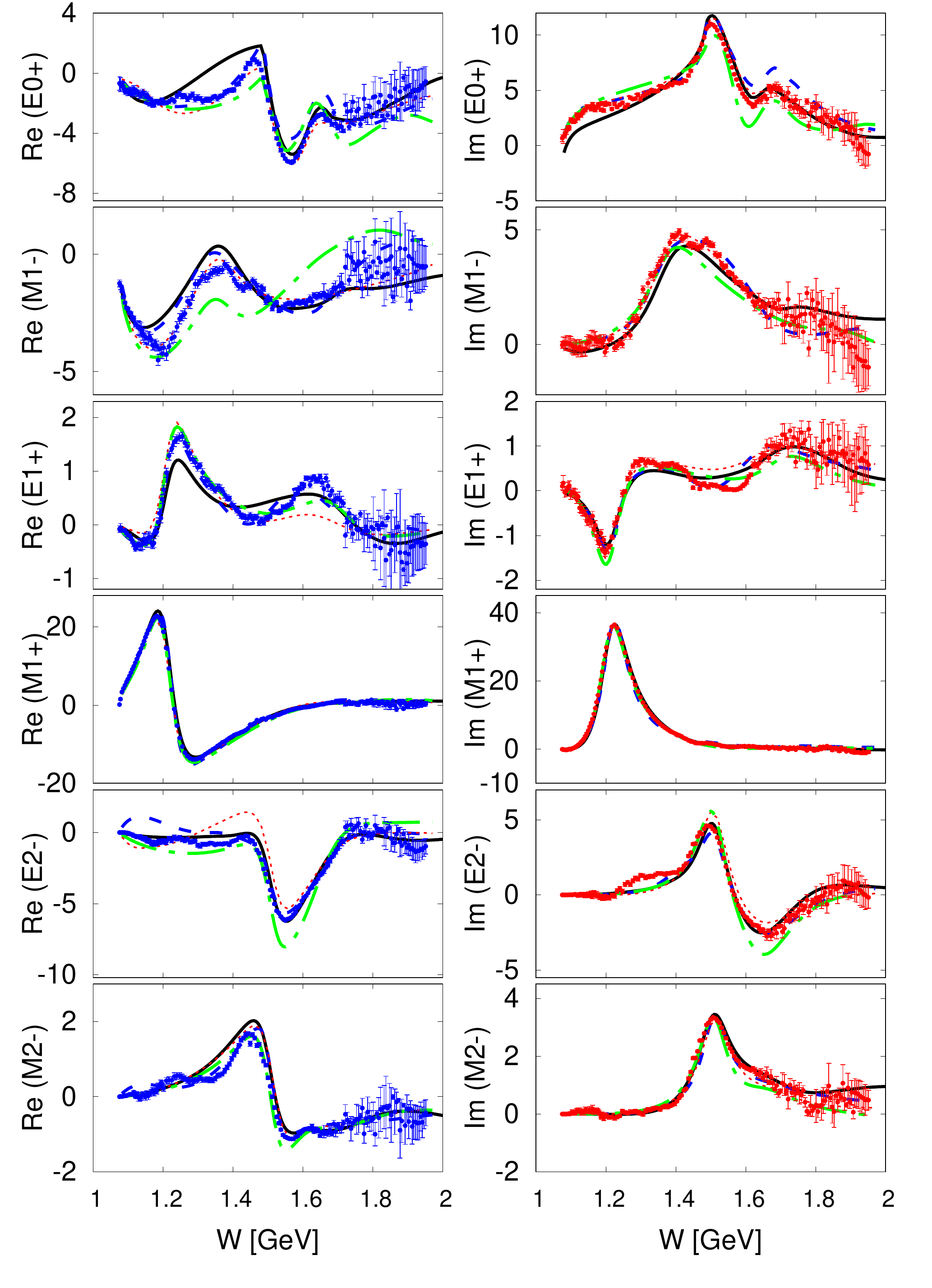}
\vspace{-5mm} \caption{\label{FigMultEDpi01+AV}Electric and magnetic multipoles from $S_{11}$, $P_{11}$, $P_{13}$, and $D_{13}$ partial waves. Real and imaginary parts of the avarage SEav solution defined in the text (blue and red dots, respectively) and
 multipoles from models (BnGa (black full lines), J\"{u}Bo (blue long-dashed lines), SAID (red short-dashed lines) and MAID2007 (green dash-dotted lines)). Multipoles are  in units of mfm.}
\end{center}
\end{figure}

In this paper we use four different models BnGa~\cite{BoGa},
J\" uBo~\cite{Juelich}, SAID~\cite{SAID}, and MAID~\cite{MAID} as initial solutions. We randomly scatter them with 30 \% uncertainty, and from these values we generate four different SE solutions, SE1, \ldots, SE4,  which form a very well defined band of solutions extending up to $W=1.7$~GeV. Comparing Figs.~\ref{FigMultSEpi01} and ~\ref{FigMultSEpi02}, it may be seen that in this energy range, our solutions demonstrate almost complete independence on initial solutions, in spite of the fact that some of them (Bonn-Gatchina for instance) has very different  $E_0+$ and $M_1-$ multipoles in the energy range \mbox{1.2 GeV  $\le W\le $ 1.4 GeV} (see Fig.~\ref{FigMultEDpi01+AV}).

Due to the fact that moderate changes in initial solutions do not cause large differences in final solutions, our method shows stability and robustness. In addition, our SE solutions are constrained by fixed-$t$ analyticity.
{ As we described in our previous work~\cite{Osmanovic2018}, our fixed-t constraint has a much deeper meaning than to ensure smoothness of our solution.
The amplitudes in our method have a well-defined crossing symmetry and possess an analytic structure postulated by the Mandelstam hypothesis.
Furthermore, it is expected that the multipoles obtained from those amplitudes also have an analytic structure as required by the Mandelstam hypothesis. }

Our SE solutions start to become poorly determined at energies $W > 1.7$~GeV.
Solutions obtained from different initial solutions differ significantly and show large errors. This is due to the lack of experimental data, especially the lack of certain polarization observables as we will further outline in the following. The number of of measured observables is much smaller. In addition, the constraining power of the fixed-$t$ analyticity decreases, especially for large $-t$ values in the backward angular range. As can be seen in Fig.~\ref{KinematicalBuund}, the unphysical $\nu$ region for large $-t$ values increases strongly. As a consequence, the Pietarinen functions in the FT AA are less constrained by experimental data and therefore,
the SE solutions are less constrained by fixed-$t$ analyticity \cite{Aznauryan}. 

\begin{figure}[h]
\begin{center}
     \includegraphics[width=0.95\textwidth]{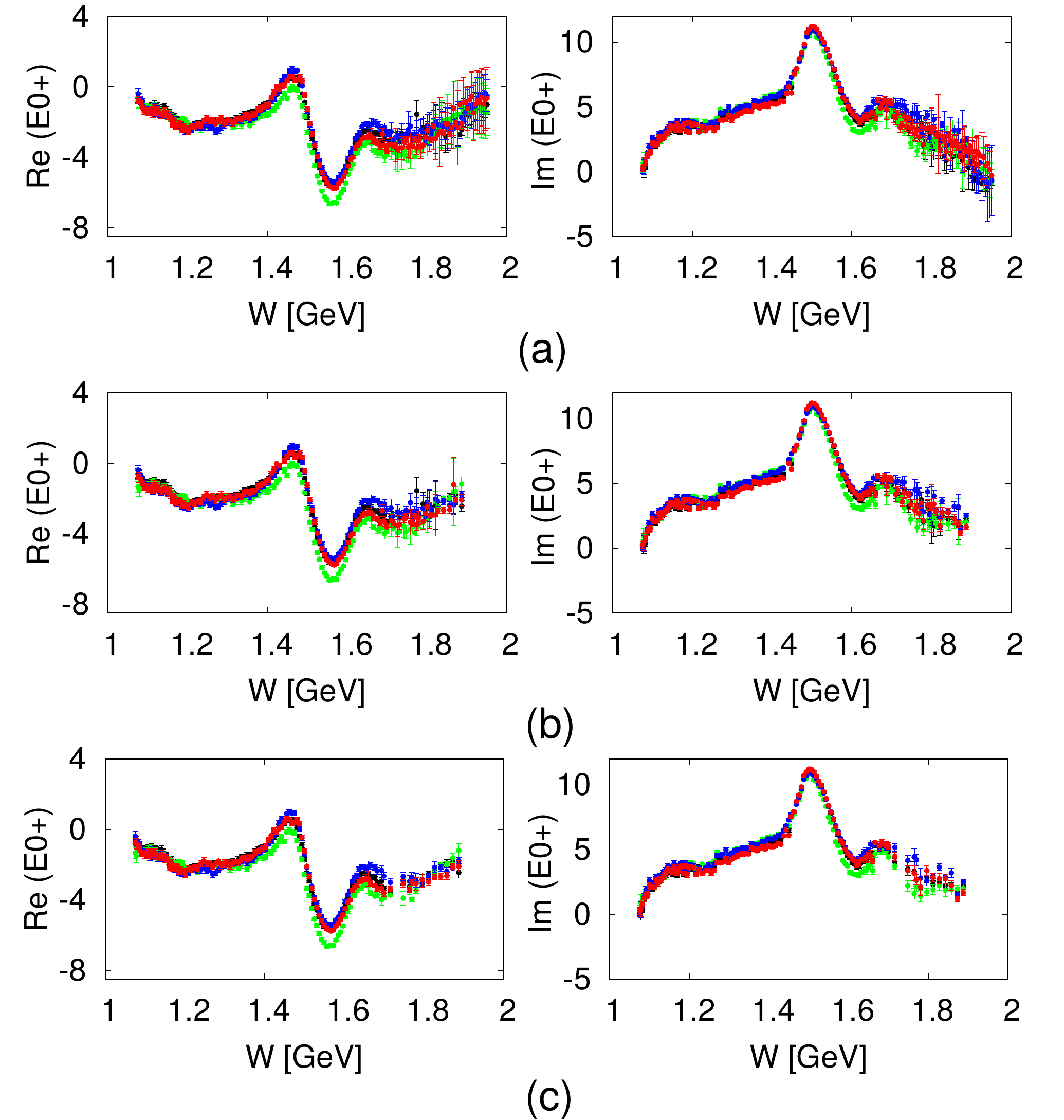}
\vspace{3mm} \caption{\label{Higherenergies} $E0+$ SE solutions obtained using different
models as initial solutions (BnGa (black), J\"{u}Bo (blue), SAID (red) and MAID (green)), with different number of observable at energies \mbox{$W > 1.7$~GeV}. In Fig.~(a) we give the standard solution, in Fig.~(b) we omit the energy points where only two observables are measured (three and more), and in  Fig.~(c) we omit the energy points where only three observables are measured (four and more). Multipoles are  in units of mfm.}
\end{center}
\end{figure}

Concerning the data, as can be seen from Table~\ref{tab:expdata}, at energies  \mbox{$W < 1.7$~GeV} we use as much as eight observables $\sigma_0$, $\Sigma$, T, P, E, F, G, H, fitting maximally five of them at the same time. However, at energies \mbox{$W > 1.7$~GeV} the number of measured observables becomes much smaller (not bigger than four, and often much less),  and the partial wave reconstruction must become less reliable. In Fig.~\ref{Higherenergies} we give an indication what happens with partial wave reconstruction for the $E0+$ multipole, when different numbers of measured observables are used in the fit. In Fig.~\ref{Higherenergies} (a) we give the solution for all energies and all measured observables, in  Fig.~\ref{Higherenergies} (b) we omit the energy points where only two observables are measured (three and more), and in  Fig.~\ref{Higherenergies} (c) we omit the energy points where only three observables are measured (four and more). It is obvious that error bars significantly drop as the number of observables grows. However, the error bars are still somewhat bigger than at lower energies, but this is entirely due to the quality of the measured data.
So, improving the data base at higher energies in the sense that both, the number of observables and the quality of measurements are increased, will improve the quality of the solution using the present technique.

\clearpage

We create the `averaged' solution by performing an average over the
four SE solutions SE1, \ldots, SE4 and taking this solution as input for the first iteration in the final fitting procedure.

\begin{figure}[htb]
\begin{center}
\includegraphics[width=0.9\textwidth]{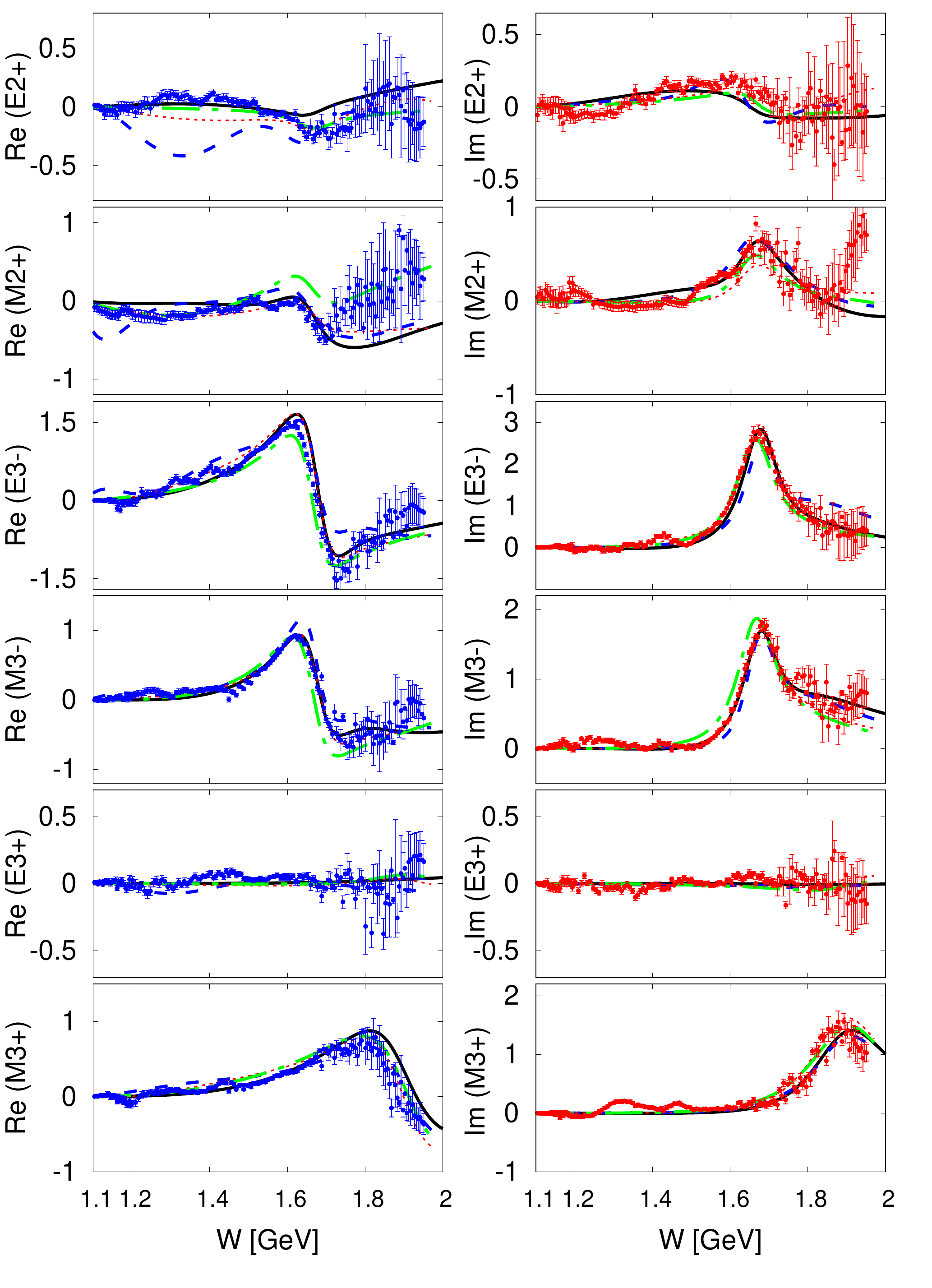}
\vspace{-5mm} \caption{\label{FigMultEDpi02+AV}Electric and magnetic multipoles from $D_{15}$, $F_{15}$, and $F_{17}$ partial waves. Real and imaginary parts of the avarage SEav solution defined in the text (blue and red dots, respectively) and
 multipoles from models (BnGa (black full lines), J\"{u}Bo (blue long-dashed lines), SAID (red short-dashed lines) and MAID2007 (green dash-dotted lines)). Multipoles are  in units of mfm.}
\end{center}
\end{figure}

%
The result for this particular solution is shown in Fig.~\ref{FigMultEDpi01+AV} and Fig.~\ref{FigMultEDpi02+AV}, where it is also compared  with all four ED solutions. The "averaged" SE solution SEav agrees fairly well with all ED solutions. As it could be expected, the SEav solution actually follows the `average' line of ED solutions. However, there are some bigger variations in the energy range 1.2 $< W < $ 1.4 GeV.



Next we will discuss what could be improved in the present data base in order to obtain SE solutions that all converge to a single one.
In spite of the abundance of experimental data in pion photoproduction, and especially for $p(\gamma,\pi^0)p$, we know that the complete set of observables needed for a unique amplitude reconstruction has still not been measured. As we see in Table~\ref{tab:expdata} a total of eight observables at various energies has already been measured. Four of them ($\sigma_0$, $\Sigma$, $T$, and $P$)  belong to the single-spin type (${\cal S}$) and the remaining four ($E, F, G, H$) belong to the beam-target type (${\cal BT}$). However, for a complete amplitude experiment (CEA), also two recoil observables out of the groups $(O_{x'}, O_{z'}, C_{x'}, C_{z'})$ of beam-recoil type ({\cal BR}) and $(T_{x'}, T_{z'}, L_{x'}, L_{z'})$ of target-recoil type (${\cal TR}$). For pion and eta photoproduction such experiments are extremely difficult as a recoil polarimetry has to be applied for the outgoing nucleon. In a pilot experiment at MAMI this has been done for (${\cal BR}$) type observables. However, for a PWA these first data points are too scarce and have large uncertainties. In the following we ignore
${\cal TR}$ type observables, which are even more difficult to measure.

\begin{figure}[htb]
\centering
      \subfloat[Single-energy fit SEav compared to the
experimental data with 4 observables at $E=0.3$~GeV ($W=1.201$~GeV)\label{subfig-1:FigAV300}]
{
     \includegraphics[width=0.65\textwidth]{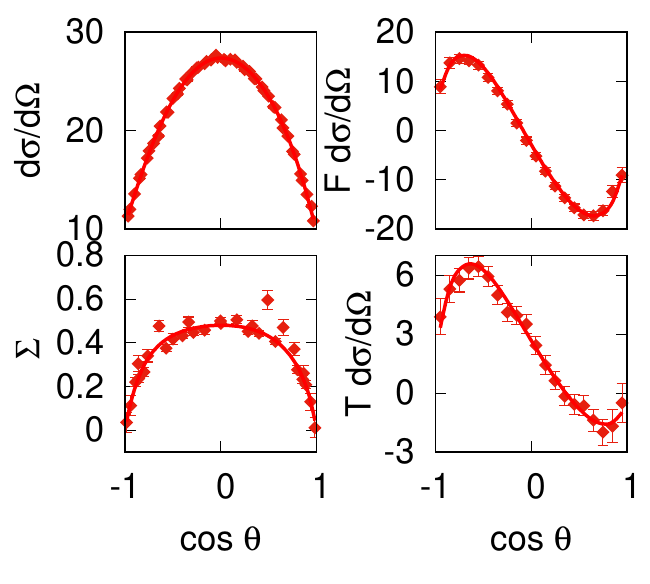}
     }
     \hfill
     \subfloat[Predictions from five different single-energy solutions, SE1 (solid black), SE2 (long-dashed blue), SE3 (short-dashed red), SE4 (dash-dotted green),
     and SEav (solid red) for polarization observables that are not
fitted $\{E, P, C_{x'}, O_{x'}\}$ (top) and $\{G, H, C_{z'}, O_{z'}\}$ (bottom) at
energie $E=0.3$~GeV ($W=1.201$~GeV). \label{subfig-2:FigPred300}]{%
       \includegraphics[width=0.95\textwidth]{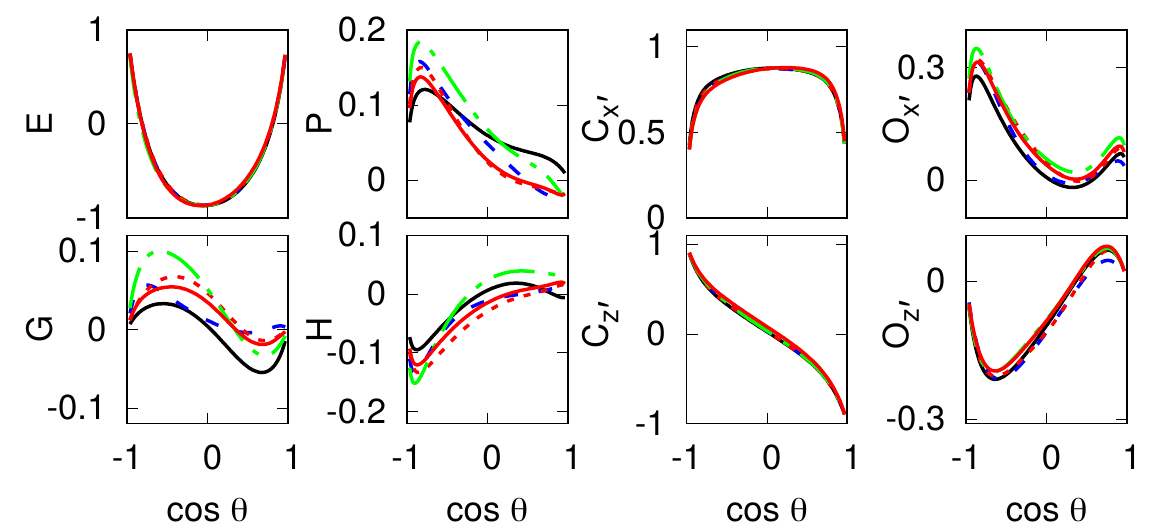}
     }
\vspace{0mm} \caption{\label{300}}
\end{figure}

\begin{figure}[htb]
\centering
      \subfloat[Single-energy fit SEav compared to the
experimental data with 5 observables at $E=0.7$~GeV ($W=1.481$~GeV)\label{subfig-1:FigAV700}]
{
     \includegraphics[width=0.85\textwidth]{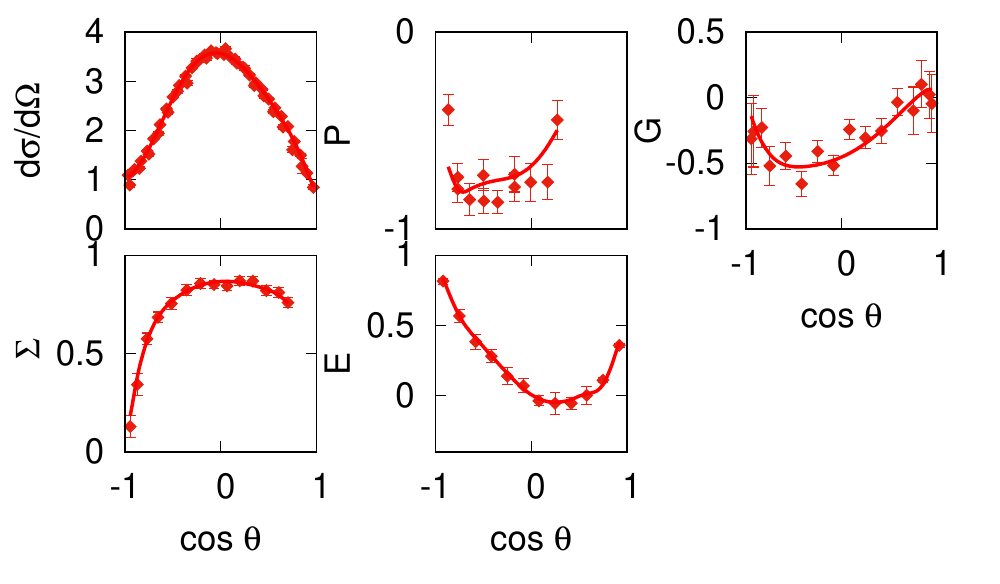}
     }
     \hfill
     \subfloat[Predictions from five different single-energy solutions, SE1 (solid black), SE2 (long-dashed blue), SE3 (short-dashed red), SE4 (dash-dotted green),
     and SEav (solid red) for polarization observables that are not
fitted $\{O_{x'}, O_{z'}, C_{x'}, C_{z'}\}$ (top) and $\{F,T,H\}$ (bottom) at
energie $E=0.7$~GeV ($W=1.481$~GeV). \label{subfig-2:FigPred700}]{%
       \includegraphics[width=0.95\textwidth]{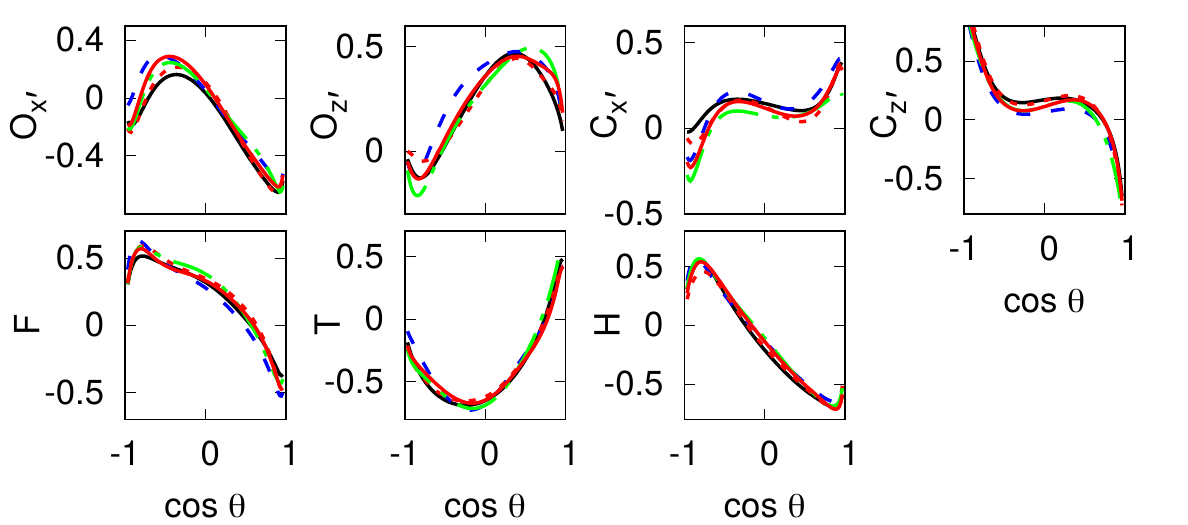}
     }
\vspace{0mm} \caption{\label{700}}
\end{figure}

\begin{figure}[htb]
\centering
      \subfloat[Single-energy fit SEav compared to the
experimental data with 4 observables at $E=1.0$~MeV ($W=1.660$~GeV)\label{subfig-1:FigAV1000}]
{
     \includegraphics[width=0.65\textwidth]{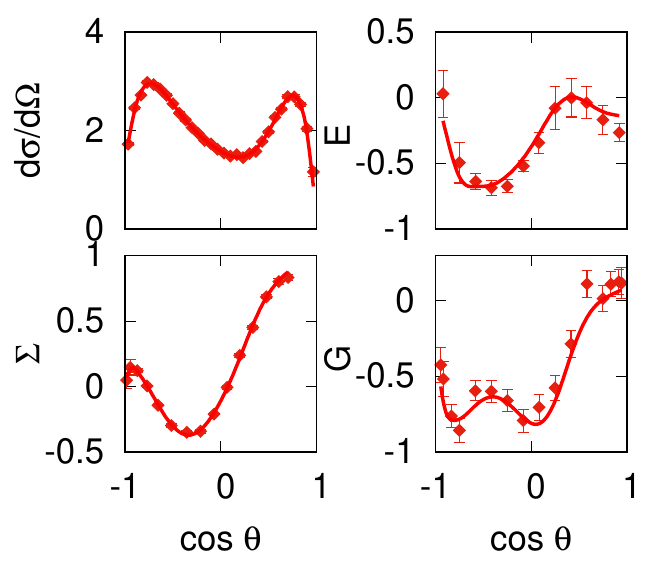}
     }
     \hfill
     \subfloat[Predictions from five different single-energy solutions, SE1 (solid black), SE2 (long-dashed blue), SE3 (short-dashed red), SE4 (dash-dotted green),
     and SEav (solid red) for polarization observables that are not
fitted $\{F, H, C_{x'}, O_{z'}\}$ (top) and $\{T, P, C_{z'}, O_{x'}\}$ (bottom) at
energie $E=1.0$~MeV ($W=1.660$~GeV).  \label{subfig-2:FigPred1000}]{%
       \includegraphics[width=0.95\textwidth]{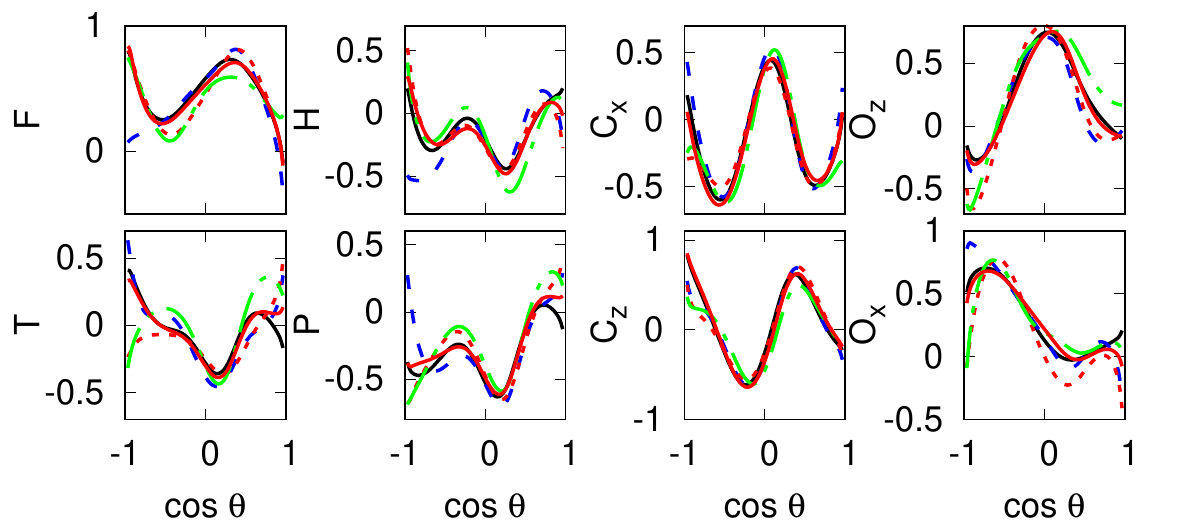}}
\vspace{0mm} \caption{\label{1000}}
\end{figure}

\begin{figure}[htb]
\centering
      \subfloat[Single-energy fit SEav compared to the
experimental data with 5 observables at $E=1.4$~GeV ($W=1.872$~GeV)\label{subfig-1:FigAV1400}]
{
     \includegraphics[width=0.85\textwidth]{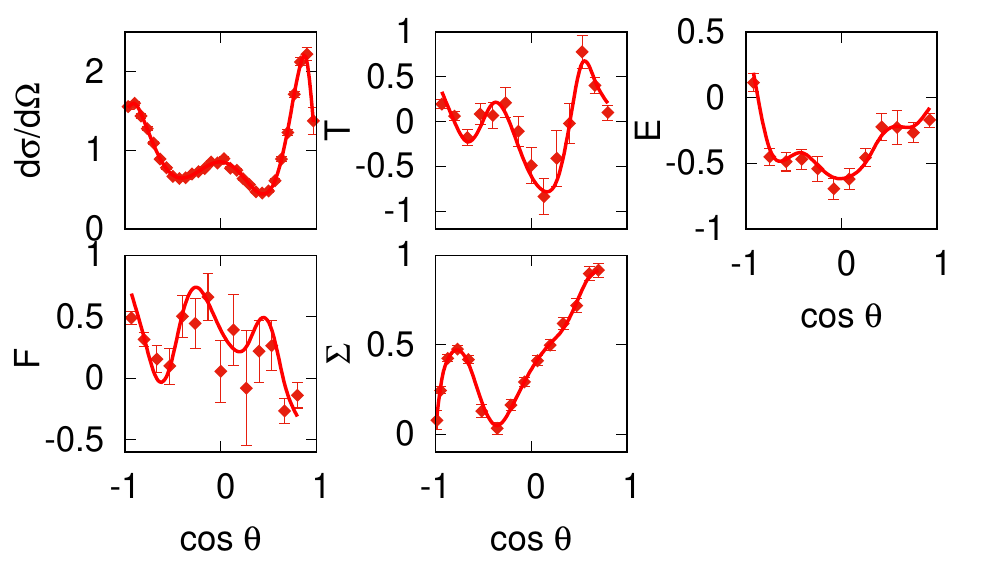}
     }
     \hfill
     \subfloat[Predictions from five different single-energy solutions, SE1 (solid black), SE2 (long-dashed blue), SE3 (short-dashed red), SE4 (dash-dotted green),
     and SEav (solid red) for polarization observables that are not
fitted $\{O_{x'}, O_{z'}, C_{x'}, C_{z'}\}$ (top) and $\{P, H, G\}$ (bottom) at
energie $E=1.4$~GeV ($W=1.872$~GeV).  \label{subfig-2:FigPred1400}]{%
       \includegraphics[width=0.95\textwidth]{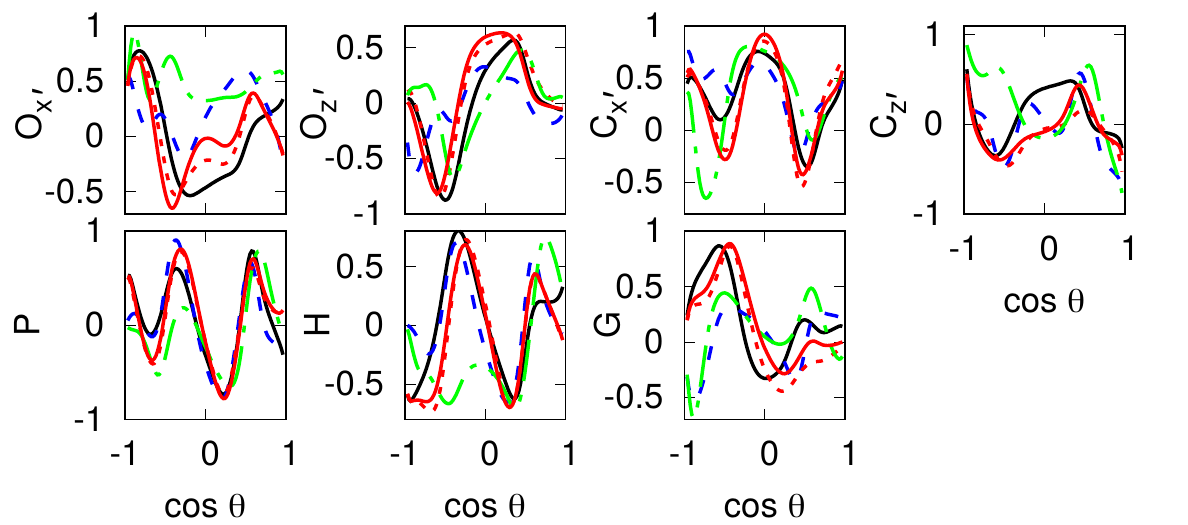}
     }
\vspace{0mm} \caption{\label{1400}}
\end{figure}

In Figs.~\ref{300} - \ref{1400} we show the results of the fit of all four SE solutions and the `averaged' SEav solution with the measured data and predictions  for a number of unmeasured observables at four representative energies $W=1.201, 1.481, 1.660,$ and $1.872$~GeV. \mbox{Figs.~\ref{300} - \ref{1400}~(a)} show the comparison of the averaged SEav solution with the measured data base at chosen energies, and it turns out that the four SE1, \ldots, SE4 solutions as well as the avaraged SEav solution give practically identical results.  Figs.~\ref{300} - \ref{1400}~(b) show the predictions of our five \mbox{SE1, \ldots, SEav} solutions for unmeasured observables which together with the measured ones can form several complete sets of observables. Different conclusions emerge for different energies.


At the lowest energy of  $W=1.201$~GeV we have four measured observables, and the measured  data are of good quality. The fit is almost perfect. Next we show the 8 unmeasured observables
out of the groups ${\cal S},{\cal BT},{\cal BR}$ predicted from our four SE1,\ldots, SE4 solutions. We can see that the agreement of all four predictions for all observables is fairly good, with the exception of $P, G$, and $H$ observables. The differences are, rather small, so only a small improvement in a PWA can be expected.


At the energy of  $W=1.481$~GeV we have five measured observables, and the measured  data are of good quality. The fit is almost perfect. Next we show the 7 unmeasured observables
out of the groups ${\cal S},{\cal BT},{\cal BR}$ predicted from our four SE1,\ldots, SE4 solutions. We see that the agreement of all four predictions for all observables is fairly good, with the exception of $O_{x'}, O_{z'}$, and $C_{x'}$ observables. The differences are, rather small, so only a little of improvement is expected.


At the energy of  $W=1.660$~GeV we have four measured observables, and the measured  data are still of good quality. The fit is almost perfect. Next we show the 8 unmeasured observables
out of the groups ${\cal S},{\cal BT},{\cal BR}$ predicted from our four SE1, \ldots, SE4 solutions. We can now see a spread for all observables at this and higher energies. The agreement is acceptable for $C_{x'},C_{z'},O_{x'}$, and $O_{z'}$, while it is notably worse for $F, H, T$, and $P$. So, we expect some improvement in the uniqueness of SE PWA if these observables are more precisely measured. We recommend to remeasure the configuration $T,P, C_{x'}, O_{z'}$ as the predictions for the beam recoil observables are fairly similar, while one must only re-measure simpler, single-spin observables $T$ and $P$.


At the highest energy of  $W=1.872$~GeV we have five measured observables, where four of them have an acceptable quality, only the $F$ observable is more uncertain. The fit is almost perfect. Next we show the 7 unmeasured observables
out of the groups ${\cal S},{\cal BT},{\cal BR}$ predicted from our four SE1, \ldots, SE4 solutions. We see that the agreement of all four predictions for all observables is not good for either of them.

\clearpage


\clearpage

\section{Summary and Conclusions}
 \label{sec:conclusions}

Using the formalism introduced and explained for $\eta$ photoproduction in Ref.~\cite{Osmanovic2018}, we have performed a fixed-$t$ single-energy partial wave analysis of $\pi^0$ photoproduction on the world collection of data.

In an iterative two-step process the single-energy multipoles are constrained by fixed-$t$ Pietarinen expansions fitted to experimental data. This leads to a partial wave expansion that obeys fixed-$t$ analyticity with a least model dependence.

In the energy range of $E=0.14 - 1.56$~GeV  ($W=1.08 - 1.95$~GeV) we have obtained electric and magnetic multipoles $E_{\ell\pm},M_{\ell\pm}$, up to $F$ waves, $\ell=3$ in 158 energy bins of about 5-10~MeV width. First we used randomized starting solutions from BnGa, J\"uBo, SAID, and MAID energy dependent solutions and obtained four
different SE solutions, SE1, \ldots, SE4 in an iterative procedure. These four SE solutions appeared already much closer together than the four underlying ED solutions, where we started from. Second we generated an `average' SE solution, SEav, again in an iterative process. All five SE solutions compare very well with the experimental data, where the `averaged' solution SEav is obtained in the least model dependent way.

Finally, we compared our five SE solutions in their predictions for unmeasured polarization observables. At lower energies the spread of these predictions is rather small, but it becomes quite large at higher energies, where it will help to propose new measurements in order to get a unique PWA.

\begin{acknowledgments}
This work was supported by the Deutsche Forschungsgemeinschaft (SFB 1044).
\end{acknowledgments}

\newpage
\section*{Appendices}
\begin{appendix}
\section{Kinematics in $\pi^0$ photoproduction}
\label{Kinematics}
For $\pi$ photoproduction on the nucleon, we consider the reaction
\begin{equation}
\gamma(k)+N(p_i)\rightarrow \pi(q)+N'(p_f)\,,
\end{equation}
where the variables in brackets denote the four-momenta of the
participating particles. In the pion-nucleon center-of-mass (c.c.) system, we define
\begin{equation}
k^{\mu}=(\omega_{\gamma},\bold{k}),\quad
q^{\mu}=(\omega_{\pi},\bold{q})
\end{equation}
for photon and pion  meson, and
\begin{equation}
p_i^\mu=(E_i,\bold{p}_i),\quad p_f^\mu=(E_f,\bold{p}_f)
\end{equation}
for incoming and outgoing nucleon, respectively. The familiar Mandelstam variables are given as
\begin{equation}
s=W^2=(p_i+k)^2,\qquad t=(q-k)^2,\qquad u=(p_f-q)^2,
\end{equation}
the sum of the Mandelstam variables is given by the sum of the external masses
\begin{equation}
s+t+u=2m_N^2+m_{\pi}^2\,,
\end{equation}
where $m_N$ and $m_{\pi}$ are masses of proton and $\pi$ meson, respectively. In the pion-nucleon center-of-mass system,  the energies and momenta can be related to the Mandelstam variable $s$ by
\begin{equation}
k=|\bold{k}|=\frac{s-m_N^2}{2\sqrt{s}},\quad
\omega=\frac{s+m_{\pi}^2-m_N^2}{2\sqrt{s}}\,,
\end{equation}
\begin{equation}
q=|\bold{q}|=\left[\left(\frac{s-m_{\pi}^2+m_N^2}{2\sqrt{s}}\right)-m_N^2\right]^{\frac{1}{2}}\,,
\end{equation}
\begin{equation}
E_i=\frac{s-m_N^2}{2\sqrt{s}},\quad
E_f=\frac{s+m_N^2+m_{\pi}^2}{2\sqrt{s}}\,,
\end{equation}
$W=\sqrt{s}$ is the c.m. energy. Furthermore, we will also refer to the lab energy of the photon, $E=(s-m_N^2)/(2m_N)$.

Starting from the $s$-channel reaction $\gamma+N\Rightarrow \pi+N$, using crossing relation, one obtains two other channels:
\begin{eqnarray}
\gamma +\pi\; &\Rightarrow&\; N+\bar{N}\qquad t-\text{channel}\,,\\
\pi + \bar{N}\; &\Rightarrow&\; \gamma + \bar{N}\qquad\;
u-\text{channel}\,.
\end{eqnarray}

All three channels defined above are described by a set of four invariant amplitudes. The singularities of the invariant amplitudes are defined by unitarity in $s$, $u$ and $t$ channels:
\begin{eqnarray}
 s-\text{channel cut:}\; &&  (m_N+m_{\pi})^2\le s <  \infty\,, \\
   u-\text{channel cut:}\; && (m_N+m_{\pi})^2\le u <  \infty\,, \\
\end{eqnarray}
and nucleon poles at $s=m_N^2$, $u=m_N^2\,$. The crossing
symmetrical variable is
\begin{equation}
\nu=\frac{s-u}{4m_N}\,.
\end{equation}

\begin{figure}[h]
\begin{center}
\includegraphics[width=7.0cm]{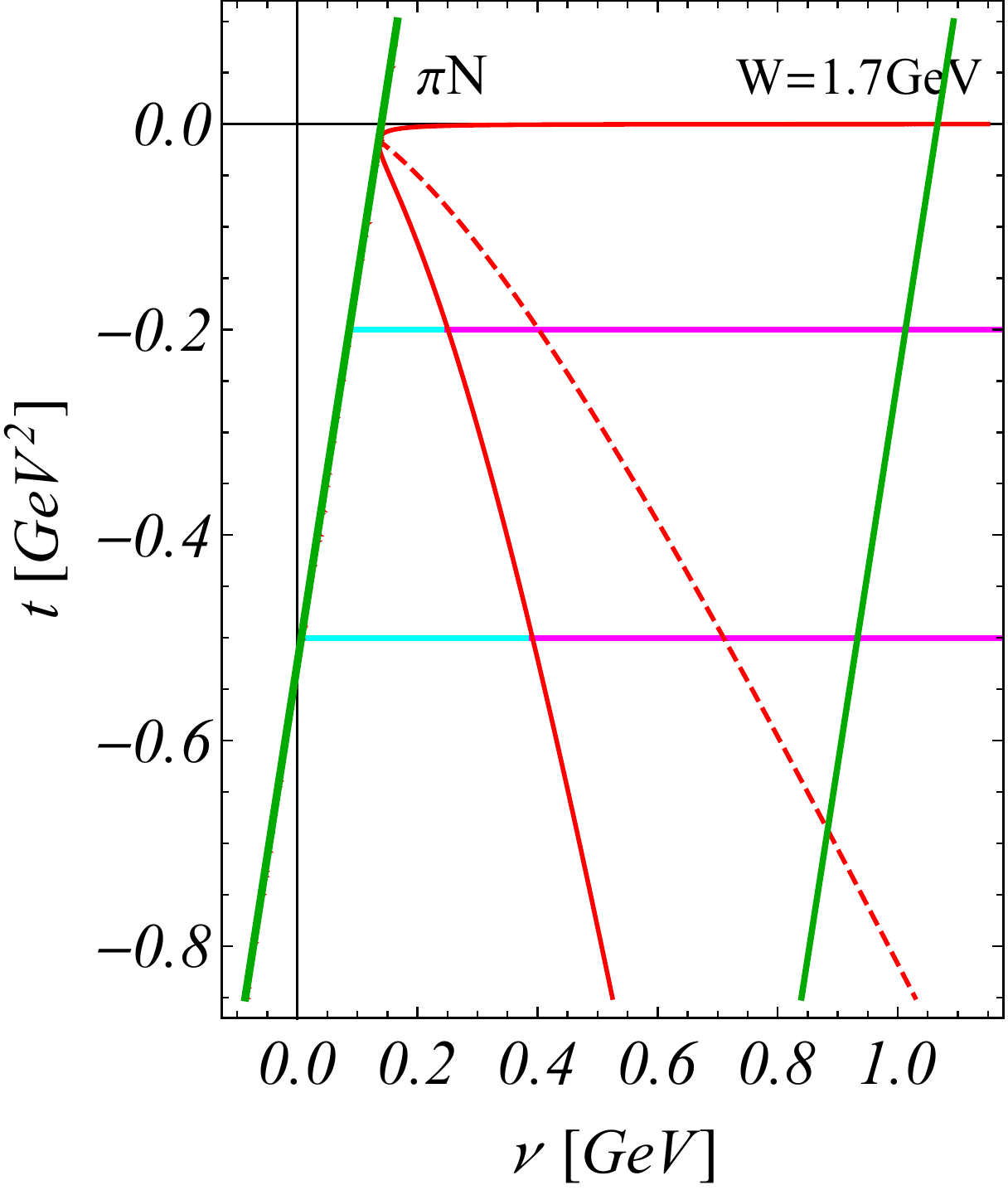}
\end{center}
\vspace{3mm} \caption{\label{KinematicalBuund}  The Mandelstam plane for pion photoproduction on the nucleon. The red solid curves are the
boundaries of the physical region from $\theta=0$ to $\theta=180^0$ and the red dashed line shows $\theta=90^0$. The green tilted vertical lines
are the threshold for pion production at W=1.073~GeV, and W=1.7~GeV. The horizontal lines denote the $t$-values $-0.2,-0.5$~GeV$^2$. The magenta parts give the
part inside the physical region, whereas the cyan parts indicate non-zero amplitudes in the unphysical region. The fixed-$t$ threshold values
for $\gamma,\pi$ in $W$ are $W_{thr}=1.208$~GeV ($t=-0.2$~GeV$^2$) and $W_{thr}=1.369$~GeV ($t=-0.5$~GeV$^2$). }
\end{figure}

The $s$-channel region is shown in Fig.~\ref{KinematicalBuund}. The upper and lower boundaries of the physical region are given by the scattering angles $\theta=0$ and $\theta=180^{\circ}$, respectively.
The c.m. energy $W$ and the c.m. scattering angle $\theta$ can be obtained from the variables $\nu$ and $t$ by
\begin{equation}
W^2=m_N(m_N+2\nu)-\frac{1}{2}(t-m_\pi^2)\,
\end{equation}
and
\begin{equation} \label{fixedt}
\mbox{cos}\,\theta\,= \frac{t-m_\pi^2 + 2\,k\,\omega}{2\,k\,q}\,.
\end{equation}
\clearpage

\section{Cross section and polarization observables}
\label{Observables}

\begin{figure}[!h]
\begin{center}
\includegraphics[width=0.5\textwidth]{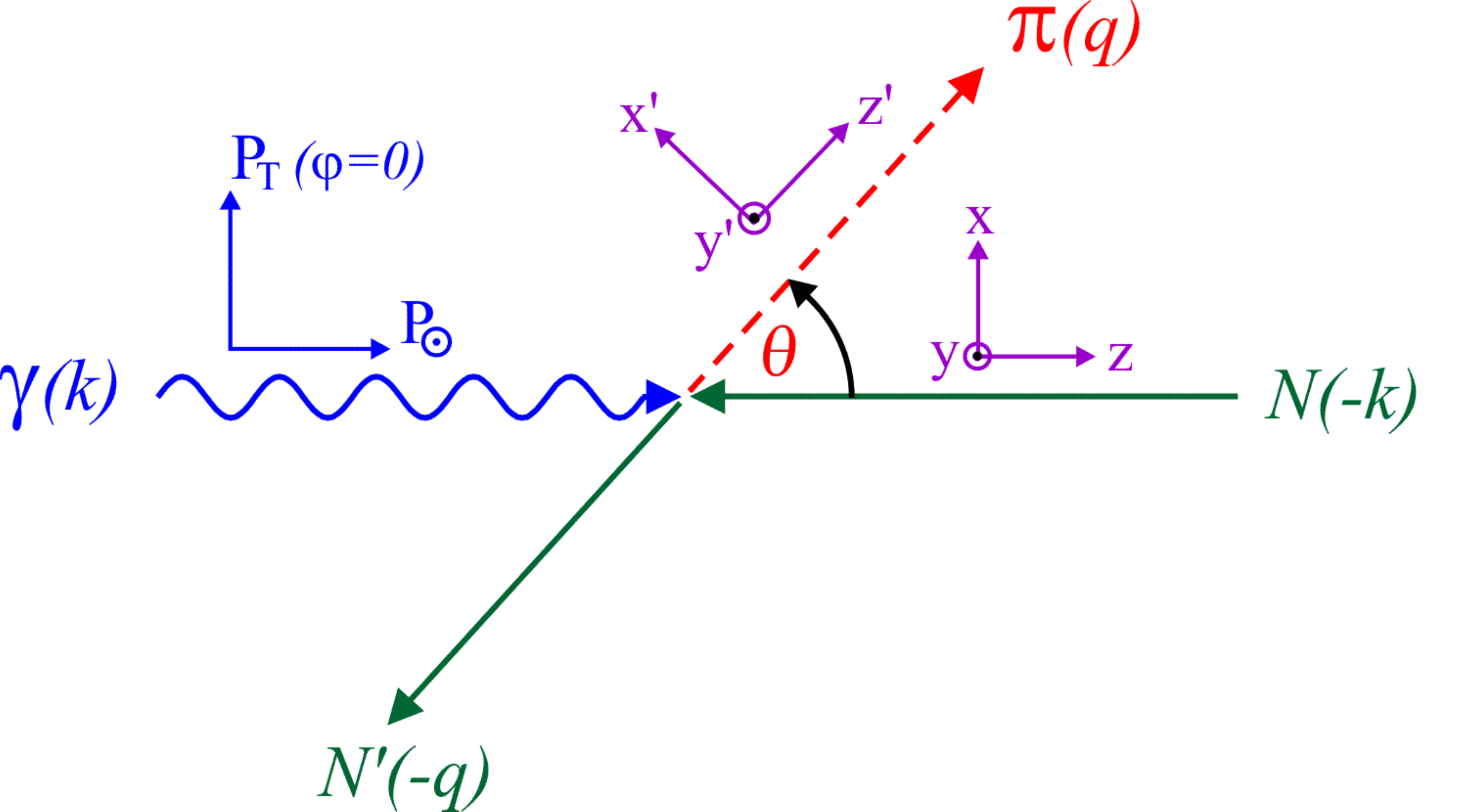} \vspace{3mm}
 \caption{ Kinematics of photoproduction and frames for polarization.
The frame $\{x,y,z\}$ is used for target polarization $\{P_x,P_y,P_z\}$, whereas the recoil polarization
$\{P_{x'},P_{y'},P_{z'}\}$ is defined in the frame $\{x',y',z'\}$, which is rotated around $y'=y$ by the polar angle $\theta$. The
azimuthal angle $\varphi$ is defined in the $\{x,y\}$ plane and is
zero in the projection shown in the figure.}\label{fig:kin}
\end{center}
\end{figure}

Experiments with three types of polarization can be performed in meson photoproduction: photon beam polarization, polarization of the target nucleon and polarization of the recoil nucleon. Target polarization will be described in the frame $\{ x, y, z \}$ in Fig.~\ref{fig:kin}, with the $z$-axis pointing into the direction of the photon momentum $\hat{ \bold{k}}$, the $y$-axis perpendicular to the reaction plane, ${\hat{\bold{y}}} = {\hat{\bold{k}}} \times {\hat{\bold{q}}} / \sin \theta$, and the $x$-axis given by ${\hat{\bold{x}}} = {\hat{\bold{y}}} \times {\hat{\bold{z}}}$. For recoil polarization we will use the frame $\{ x', y', z' \}$, with the $z'$-axis defined by the momentum vector of the outgoing meson ${\hat{\bold{q}}}$, the $y'$-axis as for target polarization and the $x'$-axis given by ${\hat{\bold{x'}}} = {\hat{\bold{y'}}} \times {\hat{\bold{z'}}}$.

The photon polarization can be linear or circular. For a linear photon polarization $(P_T=1)$ in the reaction plane $~\hat{\bold{x}}$ we get $\varphi=0$ and perpendicular, in direction ${\hat{\bold{y}}}$, the polarization angle is $\varphi=\pi/2$. For right-handed circular polarization $P_{\odot}=+1$.

We may classify the differential cross sections by the three classes of double polarization experiments and one class of triple polarization experiments, which, however, do not give additional information:
\begin{itemize}
\item polarized photons and polarized target
\end{itemize}

\begin{eqnarray}
\frac{d \sigma}{d \Omega} & = & \sigma_0
\left\{ 1 - P_T \Sigma \cos 2 \varphi \right. \nonumber \\
& & + P_x \left( - P_T H \sin 2 \varphi + P_{\odot} F \right)
\nonumber \\
& & + P_y \left( T - P_T P \cos 2 \varphi \right) \nonumber \\
& & \left. + P_z \left( P_T G \sin 2 \varphi - P_{\odot} E \right)
\right\} \, ,
\end{eqnarray}

\begin{itemize}
\item polarized photons and recoil polarization
\end{itemize}

\begin{eqnarray}
\frac{d \sigma}{d \Omega} & = & \sigma_0
\left\{ 1 - P_T \Sigma \cos 2 \varphi \right. \nonumber \\
& & + P_{x'} \left( -P_T O_{x'} \sin 2 \varphi - P_{\odot} C_{x'} \right)
\nonumber \\
& & + P_{y'} \left( P - P_T T \cos 2 \varphi \right) \nonumber \\
& & \left. + P_{z'} \left( -P_T O_{z'} \sin 2 \varphi  - P_{\odot}
C_{z'} \right) \right\} \, ,
\end{eqnarray}

\begin{itemize}
\item polarized target and recoil polarization
\end{itemize}

\begin{eqnarray}
\frac{d \sigma}{d \Omega} & = & \sigma_0 \left\{ 1 + P_{y} T + P_{y'} P
+ P_{x'} \left( P_x T_{x'} - P_{z} L_{x'} \right) \right. \nonumber \\
& & \left. + P_{y'} P_y \Sigma  + P_{z'}\left( P_x T_{z'} + P_{z}
L_{z'}\right) \right\}\,.
\end{eqnarray}

In these equations $\sigma_0$ denotes the unpolarized differential cross section, the transverse degree of photon polarization is denoted by $P_T$, $P_{\odot}$ is the right-handed circular photon polarization and $\varphi$ the azimuthal angle of the photon polarization vector in respect to the reaction plane.
Instead of asymmetries, in the following we will often discuss the product of the unpolarized cross section with the asymmetries and will use the notation $\check{\Sigma}=\sigma_0\Sigma\,, \check{T}=\sigma_0T\,,\cdots\,$. In Appendix~\ref{ObservablesCGLNandHelicity} we give expressions of the observables in terms of CGLN and helicity amplitudes.
\clearpage

\section{Observables expressed in CGLN and helicity amplitudes}
\label{ObservablesCGLNandHelicity}
Spin observables expressed in CGLN amplitudes are given by:
\begin{eqnarray}
\sigma_{0}   & = &  \,\mbox{Re}\,\left\{ \fpf{1}{1} + \fpf{2}{2} +
\sin^{2}\theta\,(\fpf{3}{3}/2
                   + \fpf{4}{4}/2 + \fpf{2}{3} + \fpf{1}{4} \right. \nonumber \\
             &   &   \mbox{} \left. + \cos\theta\,\fpf{3}{4}) - 2\cos\theta\,\fpf{1}{2} \right\} \rho \\
\check{\Sigma} & = & -\sin^{2}\theta\;\mbox{Re}\,\left\{
\left(\fpf{3}{3} +\fpf{4}{4}\right)/2
                   + \fpf{2}{3} + \fpf{1}{4} + \cos\theta\,\fpf{3}{4}\right\}\rho \\
\check{T}      & = &  \sin\theta\;\mbox{Im}\,\left\{\fpf{1}{3} -
\fpf{2}{4} + \cos\theta\,(\fpf{1}{4} - \fpf{2}{3})
                   - \sin^{2}\theta\,\fpf{3}{4}\right\}\rho \\
\check{P}      & = & -\sin\theta\;\mbox{Im}\,\left\{ 2\fpf{1}{2} +
\fpf{1}{3} - \fpf{2}{4} - \cos\theta\,(\fpf{2}{3} -\fpf{1}{4})
                   - \sin^{2}\theta\,\fpf{3}{4}\right\}\rho \\
\check{E}      & = &  \,\mbox{Re}\,\left\{ \fpf{1}{1} + \fpf{2}{2} -
2\cos\theta\,\fpf{1}{2}
                   + \sin^{2}\theta\,(\fpf{2}{3} + \fpf{1}{4}) \right\}\rho \\
\check{F}      & = &  \sin\theta\;\mbox{Re}\,\left\{\fpf{1}{3} - \fpf{2}{4} - \cos\theta\,(\fpf{2}{3} - \fpf{1}{4})\right\}\rho \\
\check{G}      & = &  \sin^{2}\theta\;\mbox{Im}\,\left\{\fpf{2}{3} + \fpf{1}{4}\right\}\rho \\
\check{H}      & = &  \sin\theta\;\mbox{Im}\,\left\{2\fpf{1}{2} +
\fpf{1}{3} - \fpf{2}{4}
                   + \cos\theta\,(\fpf{1}{4} - \fpf{2}{3})\right\}\rho \\
\check{C}_{x'} & = &  \sin\theta\;\mbox{Re}\,\left\{\fpf{1}{1} -
\fpf{2}{2} - \fpf{2}{3} + \fpf{1}{4}
                   - \cos\theta\,(\fpf{2}{4} - \fpf{1}{3})\right\}\rho \\
\check{C}_{z'} & = & \,\mbox{Re}\,\left\{2\fpf{1}{2} -
\cos\theta\,(\fpf{1}{1} + \fpf{2}{2})
                   + \sin^{2}\theta\,(\fpf{1}{3} + \fpf{2}{4})\right\}\rho \\
\check{O}_{x'} & = & \sin\theta\;\mbox{Im}\,\left\{\fpf{2}{3} - \fpf{1}{4} + \cos\theta\,(\fpf{2}{4} - \fpf{1}{3})\right\}\rho \\
\check{O}_{z'} & = & - \sin^{2}\theta\;\mbox{Im}\,\left\{\fpf{1}{3} + \fpf{2}{4}\right\}\rho\\
\check{L}_{x'} & = & - \sin\theta\;\mbox{Re}\,\left\{\fpf{1}{1} -
\fpf{2}{2} - \fpf{2}{3} + \fpf{1}{4}
                   + \sin^{2}\theta\,(\fpf{4}{4} - \fpf{3}{3})/2 \right. \nonumber \\
             &   & \mbox{} \left. + \cos\theta\,(\fpf{1}{3} - \fpf{2}{4})\right\}\rho \\
\check{L}_{z'} & = &  \,\mbox{Re}\,\left\{2\fpf{1}{2} -
\cos\theta\,(\fpf{1}{1} + \fpf{2}{2})
                   + \sin^{2}\theta\,(\fpf{1}{3} + \fpf{2}{4} + \fpf{3}{4}) \right. \nonumber \\
             &   & \mbox{} \left. + \cos\theta\sin^{2}\theta\,(\fpf{3}{3} + \fpf{4}{4})/2 \right\}\rho \\
\check{T}_{x'} & = & -\sin^{2}\theta\;\mbox{Re}\,\left\{\fpf{1}{3} +
\fpf{2}{4} + \fpf{3}{4}
                   + \cos\theta\,(\fpf{3}{3} + \fpf{4}{4})/2 \right\}\rho \\
\check{T}_{z'} & = &  \sin\theta \;\mbox{Re}\, \left\{\fpf{1}{4} -
\fpf{2}{3}
                   + \cos\theta\,(\fpf{1}{3} - \fpf{2}{4}) \right. \nonumber \\
             &   & \mbox{} \left. + \sin^{2}\theta\,(\fpf{4}{4} - \fpf{3}{3})/2 \right\}\rho \\
&& \mbox{with}\;\, \check{\Sigma}={\Sigma}\,\sigma_0\;\, \mbox{etc.
and} \;\, \rho=q/k \,.
\end{eqnarray}

The 16 polarization observables of pseudoscalar photoproduction fall into four groups, single spin with unpolarized cross section included, beam-target, beam-recoil and target-recoil observables. The simplest representation of these observables is given in terms of helicity amplitudes.

\begin{table}[ht]
\caption{Spin observables expressed by helicity amplitudes in the
notation of Barker~\cite{Barker} and Walker~\cite{Walker:1968xu}. A
phase space factor $q/k$ has been omitted in all expressions. The
differential cross section is given by $\sigma_0$ and the spin
observables $\check{O}_i$ are obtained from the spin asymmetries
$A_i$ by $\check{O}_i=A_i\,\sigma_0$.}
\begin{center}
\begin{tabular}{|c|c|c|c|}
\hline
 Observable & Helicity Representation  & Type  \\
\hline
$\sigma_0$     & $\frac{1}{2}(|H_1|^2 + |H_2|^2 + |H_3|^2 + |H_4|^2)$  &  \\
$\check{\Sigma}$ & Re$(H_1 H_4^* - H_2 H_3^*)$                           &  ${\cal S}$ \\
$\check{T}$      & Im$(H_1 H_2^* + H_3 H_4^*)$                           &   (single spin) \\
$\check{P}$      & $-$Im$(H_1 H_3^* + H_2 H_4^*)$                        &   \\
\hline
$\check{G}$      & $-$Im$(H_1 H_4^* + H_2 H_3^*)$                        &   \\
$\check{H}$      & $-$Im$(H_1 H_3^* - H_2 H_4^*)$                        &  ${\cal BT} $  \\
$\check{E}$      & $\frac{1}{2}(-|H_1|^2 + |H_2|^2 - |H_3|^2 + |H_4|^2)$ &   (beam--target)\\
$\check{F}$      & Re$(H_1 H_2^* + H_3 H_4^*)$                           &   \\
\hline
$\check{O_{x'}}$    & $-$Im$(H_1 H_2^* - H_3 H_4^*)$                        &   \\
$\check{O_{z'}}$    & Im$(H_1 H_4^* - H_2 H_3^*)$                           &  ${\cal BR}$ \\
$\check{C_{x'}}$    & $-$Re$(H_1 H_3^* + H_2 H_4^*)$                        &   (beam--recoil) \\
$\check{C_{z'}}$    & $\frac{1}{2}(-|H_1|^2 - |H_2|^2 + |H_3|^2 + |H_4|^2)$ &   \\
\hline
$\check{T_{x'}}$    & Re$(H_1 H_4^* + H_2 H_3^*)$                           &   \\
$\check{T_{z'}}$    & Re$(H_1 H_2^* - H_3 H_4^*)$                           &  ${\cal TR}$  \\
$\check{L_{x'}}$    & $-$Re$(H_1 H_3^* - H_2 H_4^*)$                        &   (target--recoil)\\
$\check{L_{z'}}$    & $\frac{1}{2}(|H_1|^2 - |H_2|^2 - |H_3|^2 + |H_4|^2)$  &   \\
\hline
\end{tabular}
\end{center}
\end{table}

\end{appendix}



\clearpage

\newpage


\begin{thebibliography}{AA}

\bibitem{BoGa} A.~V.~Anisovich {\it et al.},
Phys.\ Rev.\ C\ {\bf 96}, 055202 (2017), and references therein; and
http://pwa.hiskp.uni-bonn.de/.

\bibitem{Juelich}
D. R\"onchen, M. D\"oring, H. Haberzettl, J. Haidenbauer, U. -G. Mei\ss{}ner and K. Nakayama
Eur. Phys. J. A \textbf{51}, 70 (2015), and references therein; and
http://collaborations.fz-juelich.de/ikp/meson-baryon/main.


\bibitem{SAID}  R. L. Workman, R. A. Arndt, W. J. Briscoe, M. W. Paris, and
I. I. Strakovsky, Phys. Rev. \textbf{C 86}, 035202 (2012); and
http://gwdac.phys.gwu.edu/.

\bibitem{MAID}
  D.~Drechsel, S.~S.~Kamalov and L.~Tiator,
  Eur.\ Phys.\ J.\ A {\bf 34} (2007) 69;
and https://maid.kph.uni-mainz.de/.

\bibitem{Beck}
A.~V.~Anisovich {\it et al.},
Eur.\ Phys.\ J.\ A {\bf 52}, no. 9, 284 (2016).
\bibitem{Omelaenko:1981cr}
A.~S.~Omelaenko,
Yad.\ Fiz.\  {\bf 34}, 730 (1981).

\bibitem{Wunderlich:2017dby}
Y.~Wunderlich, A.~Švarc, R.~L.~Workman, L.~Tiator and R.~Beck,
Phys.\ Rev.\ C {\bf 96}, no. 6, 065202 (2017).

\bibitem{Workman:2016irf}
R.~L.~Workman, L.~Tiator, Y.~Wunderlich, M.~Döring and
H.~Haberzettl,
Phys.\ Rev.\ C {\bf 95}, no. 1, 015206 (2017).

\bibitem{Svarc2018}
A. \v{S}varc, Y. Wunderlich,  H. Osmanov\'{c}, M.
Had\v{z}imehmedovi\'{c}, R. Omerovi\'{c}, J. Stahov, V. Kashevarov,
K. Nikonov, M. Ostrick, L. Tiator, and R. Workman, Phys. Rev.
\textbf{C 97}, 054611 (2018).

\bibitem{Watson1954} K. M. Watson,  Phys. Rev. \textbf{95}, 228 (1954).

\bibitem{Beck:1999ge}
R.~Beck {\it et al.},
Phys.\ Rev.\ C {\bf 61}, 035204 (2000).

\bibitem{Hornidge:2012ca}
D.~Hornidge {\it et al.} [A2 and CB-TAPS Collaborations],
Phys.\ Rev.\ Lett.\  {\bf 111}, no. 6, 062004 (2013).

\bibitem{Schumann:2015ypa}
S.~Schumann {\it et al.} [A2 Collaboration at MAMI],
Phys.\ Lett.\ B {\bf 750}, 252 (2015).

\bibitem{Markou:2018skh}
L.~Markou, E.~Stiliaris and C.~N.~Papanicolas,
Eur.\ Phys.\ J.\ A {\bf 54}, no. 7, 115 (2018).

\bibitem{MAIDSE} G. Y. Chen, S. S. Kamalov, S. N. Yang, D. Drechsel, and L. Tiator,
  Phys. Rev.\textbf{ C 76}, 035206 (2007); L. Tiator, S. S. Kamalov, S. Ceci, Guan Yeu Chen, D. Drechsel,
  A. Svarc, and Shin Nan Yang, Phys. Rev. \textbf{C 82}, 055203 (2010).

\bibitem{SAIDSE} R. A. Arndt, W. J. Briscoe, I. I. Strakovsky, and
R. L. Workman, Phys. Rev. \textbf{C 74}, 045205 (2006); R. L.
Workman, R. A. Arndt, W. J. Briscoe, M. W. Paris, and I. I.
Strakovsky, Phys. Rev. \textbf{C 86}, 035202 (2012).

\bibitem{KENTSE} B. C. Hunt and D. M. Manley, Phys. Rev. C \textbf{99}, 055203 (2019)
 and Phys. Rev. C \textbf{99}, 055205 (2019).

\bibitem{Osmanovic2018}
H. Osmanovi\'{c}, M. Had\v{z}imehmedovi\'{c}, R. Omerovi\'{c}, J.
Stahov, V. Kashevarov, K. Nikonov, M. Ostrick, L. Tiator, and A.
\v{S}varc, Phys. Rev.\textbf{ C 97}, 015207 (2018).

\bibitem{HamiltonPetersen} J. Hamilton, J. L. Petersen, \emph{New development in Dispersion Theory}, Vol. 1. Nordita, Copenhagen, 1973.

\bibitem{Hohler84}
G. H\"{o}hler, \emph{Pion Nucleon Scattering}, Part 2,
Landolt-B\"ornstein: Elastic and Charge Exchange Scattering of
Elementary Particles, Vol. 9b (Springer-Verlag, Berlin, 1983).

\bibitem{Pietarinen} E. Pietarinen, Nuovo Cimento Soc. Ital. Fis. \textbf{12A}, 522 (1972).




%
%



\bibitem{Adlarson2015}
   P. Adlarson {\it et al.} [A2 Collaboration at MAMI], Phys. Rev. C {\bf 92}, 024617  (2015).

\bibitem{Leukel2001}
    R. Leukel, PhD thesis (2001) Mainz University.

\bibitem{Annand2016}
    J. R. M. Annand {\it et al.} [A2 Collaboration at MAMI] Phys. Rev. C {\bf 93}, 055209 (2016).

\bibitem{OttePhD}
P. Otte, PhD thesis (2015) Mainz University.

\bibitem{Preobrajenski2001}
I. Preobrajenski, PhD thesis (2001), Mainz University.

\bibitem{Linturi2015}
J. Linturi, PhD thesis (2015) Mainz University.

\bibitem{Ahrends2005}
J. Ahrens {\it et al.} Eur. Phys. J. A {\bf 26}, 135 (2005).

\bibitem{Hartmann2014}
   J. Hartmann {\it et al.} [CBELSA/TAPS Collaboration] Phys. Rev. Lett. {\bf 113}, 062001 (2014).

\bibitem{Gottschall2014}
  M. Gottschall {\it et al.} [CBELSA/TAPS Collaboration] Phys. Rev. Lett. {\bf 112}, 012003 (2014).

\bibitem{Thiel2012}
    A. Thiel {\it et al.} [CBELSA/TAPS Collaboration] Phys. Rev. Lett. {\bf 109}, 102001 (2012).

\bibitem{Bartalini2005}
O. Bartalini {\it at. al.} Eur. Phys. J. A {\bf 26}, 399  (2005).

\bibitem{deBoor}
C. de Boor, \emph{A Practical Guide to Splines}, Springer-Verlag,
Heidelberg, 1978, revised 2001.

\bibitem{Aznauryan} I. G. Aznauryan,    Phys. Rev. C {\bf 67}, 015209 (2003).
\bibitem{Pasquini} B. Pasquini, D. Drechsel, and L. Tiator, Eur. Phys. J. A {\bf 27}, 231
(2006).
\bibitem{Barker}I. S. Barker, A. Donnachie, J. K. Storrow,
Nucl. Phys. B {\bf 95}, 347 (1975).

\bibitem{Walker:1968xu}
  R.~L.~Walker,
  Phys.\ Rev.\  {\bf 182}, 1729 (1969).

\end{thebibliography}
\end{document}